\documentclass[11pt,a4paper]{article}
\pdfoutput=1

\usepackage{epsfig}
\usepackage{latexsym}
\usepackage{amsfonts}
\usepackage{amsmath}
\usepackage{amsthm}
\usepackage{amssymb}
\usepackage{amsbsy}
\usepackage{multirow}
\usepackage{slashed}
\usepackage{color}
\usepackage{mathrsfs}
\usepackage{subfigure}
\usepackage[font={footnotesize,sl},bf]{caption}
\usepackage{wrapfig}

\usepackage{jheppub}

\def\be{\begin{equation}}
\def\ee{\end{equation}}
\def\bea{\begin{eqnarray}}
\def\eea{\end{eqnarray}}

\newcommand{\nn}{\nonumber \\}

\title{Multi-centered D1-D5 solutions at finite B-moduli}
\author{Borun D.\ Chowdhury,}
\author{Daniel R.\ Mayerson}
\affiliation{Institute for Theoretical Physics, University of Amsterdam,\\
Science Park 904, Postbus 94485, 1090 GL Amsterdam, The Netherlands}

\emailAdd{b.d.chowdhury@uva.nl,d.r.mayerson@uva.nl}
\abstract{We study the fate of two-centered D1-D5 systems on $T^4$ away from the singular supergravity point in the moduli space. We do this by considering a background D1-D5 black hole with a self-dual B-field moduli turned on and treating the second center in the probe limit in this background. We find that in general marginal bound states at zero moduli become metastable at finite B-moduli, demonstrating a breaking of supersymmetry. However, we also find evidence that when the charges of both centers are comparable, the effects of supersymmetry breaking become negligible. We show that this effect is independent of string coupling and thus it should be possible to reproduce this in the CFT at weak coupling. We comment on the implications for the fuzzball proposal.
}
\keywords{Black Holes in String Theory, AdS/CFT Correspondence, Black Holes}

\begin{document}
\maketitle

\section{Introduction}

\subsection*{D1-D5 system and moduli}
The D1-D5 system has been very useful for studying black holes and black strings in the D1-D5 system. This system flows in the IR to a $1+1$ dimensional $\mathcal N=(4,4)$ CFT~\cite{Larsen:1999uk} making this system suitable for holographic studies.
A sample of various interesting results obtained using this system is (i) matching of the Bekenstein-Hawking entropy with the entropy of the dual CFT~\cite{Strominger:1996sh}, (ii) reproduction of the Hawking radiation from a certain irrelevant deformation of the dual CFT~\cite{Callan:1996dv,Dhar:1996vu,Das:1996wn,Das:1996jy,Maldacena:1996ix,Mathur:1997et,Gubser:1997qr}, (iii) explicit constructions of the microstates of the two charge system (two charge fuzzballs)~\cite{Lunin:2002qf,Lunin:2002iz,Taylor:2005db,Skenderis:2006ah,Kanitscheider:2007wq} and matching their entropy computed using geometric quantization with that of two charge solutions in the dual CFT~\cite{Rychkov:2005ji}, (iv) construction of many three-charge supersymmetric microstates of the system in the bulk (for reviews see~\cite{Mathur:2005zp,Bena:2004de,Skenderis:2008qn,Balasubramanian:2008da,Chowdhury:2010ct}, (v) construction of a family of non-extremal smooth geometric solutions known as JMaRT~\cite{Jejjala:2005yu} and (vi) the identification of the classical instability of JMaRT solutions with their Bose-enhanced Hawking radiation~\cite{Chowdhury:2007jx,Avery:2009tu}.

Many of the results mentioned above involve a matching of results obtained from supergravity and those obtained from the CFT at the  ``orbifold point"~\cite{Larsen:1999uk,David:2002wn}. Not only are these different points in the moduli space, but the supergravity point\footnote{More accurately this is a co-dimension four manifold and not a point for the case of $S^1 \times T^4$ compactification~\cite{Larsen:1999uk}.} is even singular. This singularity comes from the D1 and D5 branes being marginally bound and thus continuously separable. This makes the dual CFT singular~\cite{Seiberg:1999xz}. This makes some of the results above puzzling, if not suspicious, and motivates one to move away from these special points in the moduli space to understand the system better. Perturbative deviations from the orbifold point were studied in~\cite{Gava:2002xb,David:2008yk} and developed further in~\cite{Avery:2010er,Avery:2010hs,Avery:2010vk,Burrington:2012yq} and these techniques were used to study evolution of entanglement entropy~\cite{Asplund:2011cq}. Techniques to move away from the singular supergravity point by turning on certain moduli (self-dual $B_{NS}$-field on the compact $T^4$ which the D5 branes wrap) were studied in~\cite{Maldacena:1999mh,Dhar:1999ax,Lunin:2003gw,Lee:2008ha,Breckenridge:1996tt,Costa:1996zd,Costa:1997zy}. In this paper we will study the effect of turning on the said self-dual B-moduli on multi-centered configurations and will refer to this moduli simply as the moduli.

\subsection*{Entropy enigma, moduli and lifting of long multiplets}

In~\cite{Bena:2011zw,Chowdhury:2012ff}  a family of novel supersymmetric phases of the D1-D5 CFT, which in certain ranges of charges have more entropy than all known ensembles, was found at the orbifold point. Further,  bulk BPS configurations that exist in the same range of parameters as these phases, and have more entropy than a BMPV black hole were also found.  These configurations are the first  instance of black hole entropy enigma~\cite{Denef:2007vg,Denef:2007yn} with a controlled CFT dual. The entropy of the bulk configurations is smaller than that of the CFT phases, which indicates that some of the CFT states are lifted at strong coupling. Neither the bulk nor the boundary phases are captured by the elliptic genus, which makes the coincidence of the phase boundaries particularly remarkable.

According to common lore, at strong coupling all long multiplets become heavy and only short multiplets contribute to the partition function. The elliptic genus counts only short multiplets and equals the entropy of the single-center BMPV black hole. The supersymmetric multi-centered configurations of~\cite{Bena:2011zw,Chowdhury:2012ff}, carrying more entropy than the BMPV black hole, thus do not seem to have a natural place in this story. Ref.~\cite{Dabholkar:2009dq} investigated why multi-centered dyonic configurations, where at least one of the centers is a quarter-BPS black hole, do not contribute to the index which counts quarter-BPS multiplets in $\mathcal N=4, D=4$ supergravity. They found that  such multi-centered configurations are continuously connected to long multiplets and thus do not contribute to the index.

Following~\cite{Dabholkar:2009dq}, one would expect that the configurations in~\cite{Bena:2011zw,Chowdhury:2012ff} are also continuously connected to long multiplets and would become non-supersymmetric after turning on some moduli which takes them away from a supersymmetric sub-manifold. However, this leaves one with the question of what happens to this multi-centered solutions -- are they discontinuous and suddenly disappear or do they become slightly non BPS? A further  question is what happens to the Penrose process like instability studied in~\cite{Chowdhury:2011qu,Anninos:2011vn} under such changes in moduli. In fact, one may also wonder what happens to all the multi-centered configurations discussed in~\cite{Bena:2004de}. While the answer to the last question is more involved since it involves fully backreacted multi-centered solutions, we initiate a study in this direction by answering the first two questions by looking at a black hole solution with probe supertubes after turning on a component of the self-dual $B_{NS}$-moduli on the compact $T^4$ which the D5-branes wrap.

\subsection*{A simple argument showing lifting of multi-centered configurations}

\begin{figure}[htbp]
\begin{center}
\subfigure[]{
\includegraphics[scale=.25]{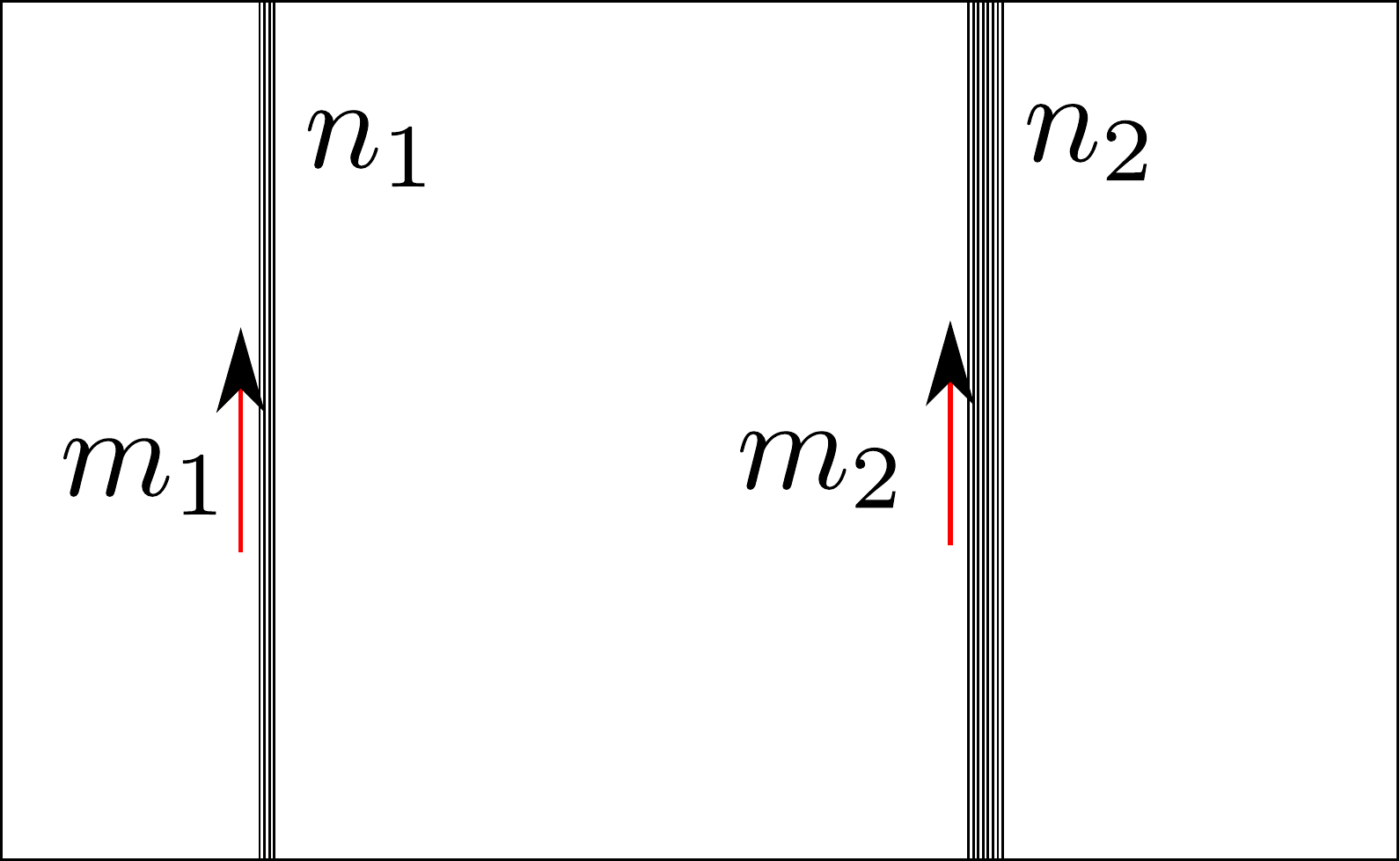}} \hspace{1in}
\subfigure[]{
\includegraphics[scale=.25]{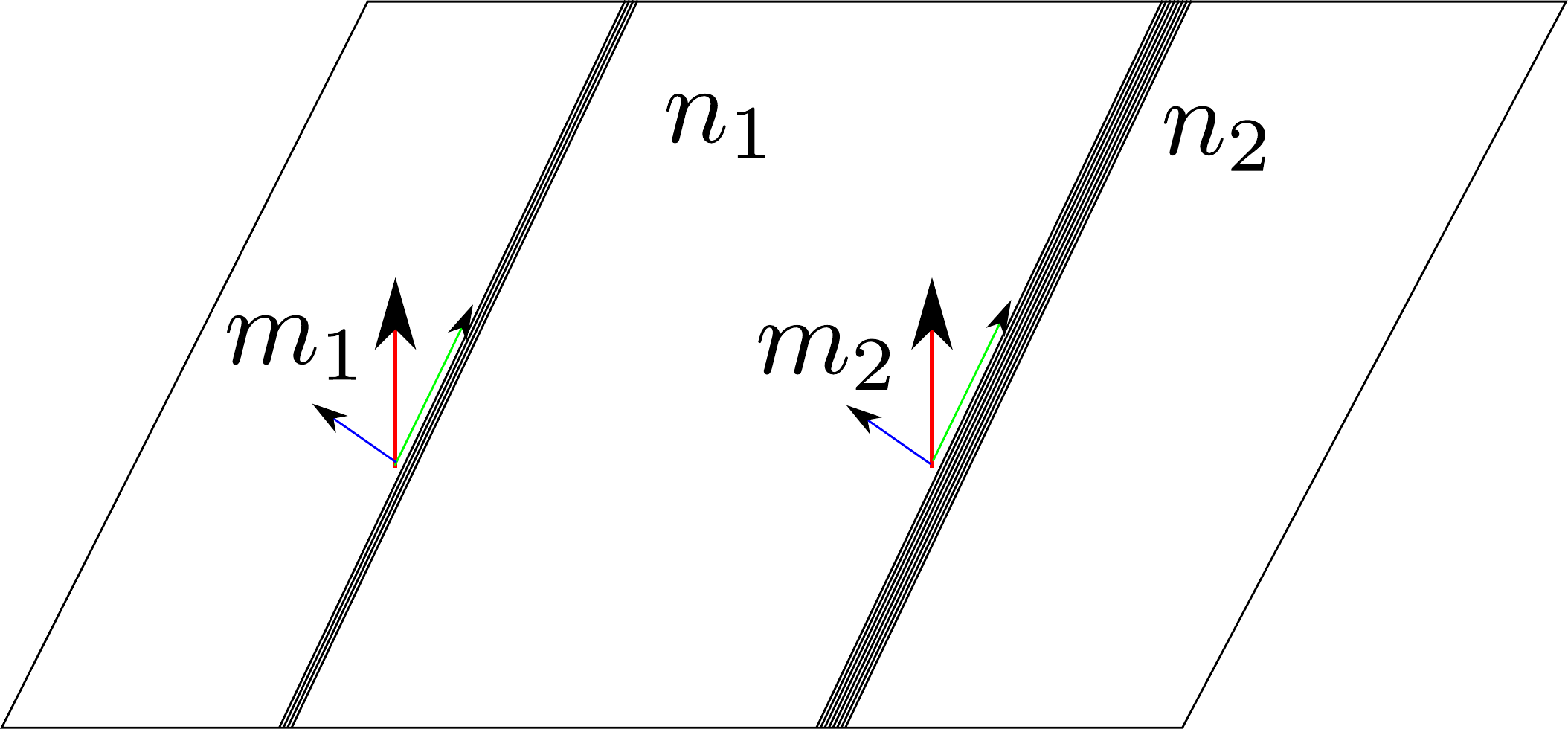}}
\caption{In (a) a rectangular torus is shown with the momentum and windings at the two centers. The charges are aligned along the sides of the torus. This configuration is supersymmetric. In (b) the torus is tilted. The winding runs along the lattice while the momentum runs along the reciprocal lattice. This means that for the same charges as in (a), the winding and momentum are not aligned. The momentum perpendicular to a stack of branes is interpreted as giving it velocity. Unless the velocities of the two stacks is the same they are not supersymmetric. For a finite size torus, quantization of momentum ensures that the velocities are not the same in general.
}
\label{weakCoupling}
\end{center}
\end{figure}

While we present a detailed analysis of our results in the bulk of the paper, here we present a simple argument to show why multi-centered solutions become non-supersymmetric when turning on generic values of a certain modulus. Consider type IIA (or IIB) compactified on $T^2$. Let the lattice vectors of the $T^2$ be given by  $\partial_x$ and $\partial_y$ and the reciprocal lattice vectors perpendicular to these be given by $\partial^\perp_x$ and $\partial^\perp_y$ respectively.  Consider the first center consisting of $n_1$ fundamental strings running along $\partial_y$ and $m_1$ units of momentum along $\partial^\perp_x$. Similarly, the other center consists of $n_2$ fundamental strings and $m_2$ units of momentum along the same directions as the first center. When the torus is rectangular the momentum at each center is aligned with the strings and, as is well known, two such centers at rest with respect to each other are supersymmetric. Such a configuration is shown in figure \ref{weakCoupling}a. However, when the torus is not rectangular, the momentum at each center is not aligned with the winding. The momentum can then be resolved in directions parallel and perpendicular to the windings as shown in figure \ref{weakCoupling}b. The bound state of a string (with or without momentum parallel to it) with momentum perpendicular to it ($p_\perp$) is the same string moving with a velocity which accounts for $p_\perp$. The two centers in figure  \ref{weakCoupling}b are thus interpreted as having velocities perpendicular to the windings and will be mutually supersymmetric {\em only if} the velocities are the same (such that they are at rest with respect to each other). Because of the quantization of momentum on a finite size torus this is not possible for generic charges and we see that for generic values of the modulus, two-centered configurations cannot be mutually supersymmetric.

\textbf{Remark:\footnote{We wish to thank the anonymous referee for bringing this to our attention and especially Sameer Murthy for illuminating discussions on this subject.}} 
The F1/P system considered here consists of two 1/2 BPS centers (in $\mathcal{N}=4$ language) and according to the discussion in~\cite{Dabholkar:2009dq} one would naively expect it to contribute to the elliptic genus by being supersymmetric at all values of moduli, although we have argued otherwise. This brings us to an important subtlety. Two {\em truly} bound centers rest at an inter-center separation which comes from balancing of attractive and repulsive forces. The equation which gives this separation is known also as the bubble equation. The claim of~\cite{Dabholkar:2009dq} is that when both centres are 1/2 BPS the inter-center separation readjusts with moduli to maintain SUSY and so such configurations contribute to the elliptic genus. However, when one or both the centres are 1/4 BPS, no readjustment can always maintain SUSY everywhere in the moduli space.
 Two F1-P centres are {\em marginally bound} in that they have neither attractive nor repulsive forces at zero moduli and so can be at any separation from each other. When moduli are turned on they become non-supersymmetric as we have shown and this results in them becoming attractive. The true inter-center separation in this case is zero and at that point they should be thought at a single center 1/2 BPS configuration and they contribute to the elliptic genus as single center configurations.
Nevertheless our example serves to show the mechanism of susy-breaking we consider in a nice intuitive way.

\subsection*{Overview of the paper}

The plan of the rest of the paper is as follows. In section \ref{Section:D1-D5-System} we review the procedure of~\cite{Maldacena:1999mh,Dhar:1999ax,Lunin:2003gw} to turn on self-dual B field on the compact $T^4$ in the D1-D5-P system. We also introduce the probe supertube in this background, generalizing methods of e.g.~\cite{Chowdhury:2011qu}. In section \ref{Section:ExtremalBHs} we analyze (numerically) the probe supertube in a BMPV black hole background; we see that supersymmetry is broken for generic multi-centered solutions because of an analogous charge quantization argument as given above in the F1-P system. However, we also give evidence that the breaking of supersymmetry would be invisible in supergravity when both centers have large and comparable charges (i.e. in the fully backreacted geometry). In section \ref{Section:NonExtremalBHs} we generalize to the study of probe tubes in non-extremal black hole backgrounds to extend the analysis of~\cite{Chowdhury:2011qu} at finite B-moduli. We end with a summary of our results and conclude in section \ref{Section:Conclusion}.

\section{D1-D5 system with B-moduli} \label{Section:D1-D5-System}
We are interested in studying a multi-centered configuration of the D1-D5 system with the self-dual $B_{NS}$ moduli on the compact $T^4$ turned on. This ensures that the system moves away from the Seiberg-Witten point in the moduli space~\cite{Seiberg:1999xz} where the CFT is singular. We do this by following the procedure that was used in~\cite{Maldacena:1999mh,Dhar:1999ax,Lunin:2003gw,Lee:2008ha} to turn on these moduli. We will concentrate on a two-centered configuration consisting of a non-extremal D1-D5-P rotating black hole and a supertube. It would be interesting to study the system where the supertube is fully backreacted; we leave that for future work. For now, we will focus on the supertube in the probe limit as was done in~\cite{Marolf:2005cx,Bena:2008dw,Chowdhury:2011qu}. Thus we will follow the procedure of~\cite{Maldacena:1999mh,Dhar:1999ax,Lunin:2003gw,Lee:2008ha} to turn on moduli on a non-extremal black hole and then probe it with a supertube. We can then obtain the extremal black hole geometry with finite B-moduli by simply taking the extremal limit of this non-extremal geometry.

\subsection{Initial D1-D5-P Black Hole}\label{sec:modinitial}
Our starting point for the moduli-adding procedure is the non-extremal rotating D1-D5-P black hole~\cite{Cvetic:1996xz,Cvetic:1996kv,Giusto:2004id}, where five dimensions parametrized by $z,y_1,y_2,y_3,y_4$ are compactified into a $T^5$ and the other five parametrized by $t,r,\theta,\phi,\psi$ are not. There are six parameters that define the solution: a mass parameter $m$, three dimensionless numbers $\delta_i$ related to the charges, and two angular momentum parameters $a_1, a_2$. We denote:
\begin{equation}
 c_i \equiv \cosh\delta_i, \qquad s_i \equiv \sinh \delta_i.
\end{equation}
The D1-D5-P system (before turning on moduli) is then given by:
\begin{align}
\label{mod:metricinitial}
ds^2 &= ds^2_6+ \frac{H_1^{1/2}}{H_2^{1/2}}\left[dy_1^2+dy_2^2+dy_3^2+dy_4^2\right],\\
e^{2\phi} &= \frac{H_1}{H_2},\\
 C_2 &= K_1,\\
 C_6 &= -K_2\wedge dy_1\wedge dy_2\wedge dy_3\wedge dy_4,\\
\label{mod:B2C4initial} B_2 &= C_4=0,
\end{align}
where we have used the two-forms $K_1, K_2$ and the six-dimensional metric element $ds_6^2$, defined as:
\begin{align}
 K_1 & = \frac{c_1}{s_1} dt\wedge dz- \frac{c_1}{s_1} H_1^{-1} (dt+k)\wedge dz -B_1\wedge dz -\frac{c_1 c_3}{s_1 s_3} H_1^{-1}dt\wedge dk\nn
 & + m s_2 c_2 \frac{r^2+a_2^2+m s_1^2}{f H_1}\cos^2\theta d\psi\wedge d\phi - \frac{s_3}{c_3} dt\wedge B_1 - \frac{c_1}{s_1} H_1^{-1} dt\wedge B_3,\\
K_2 &= K_1 \textrm{ with } \left(s_1\leftrightarrow s_2; c_1\leftrightarrow c_2; H_1\leftrightarrow H_2\right),\\
ds_6^2 & = \frac{1}{H_3 (H_1 H_2)^{1/2}}\left[-H_m\left(dt+k\right)^2 + \left( \frac{c_3}{s_3}(1-H_3) dt + \frac{c_3}{s_3} k + H_3 B_3 + H_3 dz\right)^2\right] + (H_1 H_2)^{1/2}ds_4^2 \\
ds_4^2 &= f \left( \frac{r^2}{g} dr^2 +  d\theta^2+\sin^2\theta d\phi^2+\cos^2\theta d\psi^2\right)\nn
&+ H_m^{-1} \left(a_1\cos^2\theta d\psi + a_2\sin^2\theta d\phi\right)^2 - \left(a_2\cos^2\theta d\psi + a_1\sin^2\theta d\phi\right)^2,\\
k & = \frac{m}{f}\left[ -\frac{c_1 c_2 c_3}{H_m}\left(a_1\cos^2\theta d\psi + a_2\sin^2\theta d\phi\right) + s_1 s_2 s_3\left(a_2\cos^2\theta d\psi + a_1\sin^2\theta d\phi\right)\right],\\
B_i & = \frac{m}{f H_m} \frac{c_j c_k}{s_i} \left(a_1\cos^2\theta d\psi + a_2\sin^2\theta d\phi\right).
\end{align}
Everything is built from the following functions:
\begin{align}
 H_i &= 1 + \frac{m s_i^2}{f},&  H_m &= 1 - \frac{m}{f},\\
f &= r^2 + a_1\sin^2\theta + a_2^2\cos^2\theta,& g &= (r^2+a_1^2)(r^2+a_2^2)-mr^2 = (r^2-r_+^2)(r^2-r_-^2),
\end{align}
where the inner and outer horizons are given by:
\be \label{mod:horizonsinitial} (r_{\pm})^2 = \frac12\left(m-a_1^2-a_2^2\pm\sqrt{\left(m-a_1^2-a_2^2\right)^2-4a_1^2a_2^2}\right) . \ee
The ADM mass, electric charges, and angular momenta of the 5D black hole are (in units where $G_5=\pi/4$, see appendix \ref{sec:appunits}):
\begin{align}
M_{ADM} &= \frac{m}{2}\sum_i \left(s_i^2+c_i^2\right),& J_{\psi} &= -m(a_1 c_1 c_2 c_3 - a_2 s_1 s_2 s_3),\\
\label{mod:initialQi} Q_i & = m s_i c_i,& J_{\phi} &= -m(a_2 c_1 c_2 c_3 - a_1 s_1 s_2 s_3).
\end{align}
The supersymmetric extremal limit\footnote{There is also a non-supersymmetric extremal limit, obtained by putting $m=(|a_1|+|a_2|)^2$, for which $M_{ADM} > \sum_i Q_i$. This is the 'ergo-cold' black hole studied in~\cite{Dias:2007nj}.} is obtained by taking the limit $m,a_1,a_2\rightarrow 0$ together with $|\delta_i|\rightarrow\infty$ while keeping the charges $Q_i$ and the ratios $a_i/\sqrt{m}$ fixed. One recovers the supersymmetric rotating BMPV black hole~\cite{Breckenridge:1996is} with $M_{ADM} = \sum_i Q_i$ and $|J_{\phi}|=|J_{\psi}|$. Note that for the extremal black hole, the horizon is at $r_+=0$, and the functions $H_i$ are $H_i = 1 + |Q_i|/r^2$.

\subsection{D1-D5-P Black Hole With Moduli}\label{sec:D1D5wmod}
To turn on the moduli for the D1-D5-P system, we follow the procedure of~\cite{Maldacena:1999mh,Dhar:1999ax,Lunin:2003gw}. This consists of first doing two T-dualities along the $y_3$ and $y_4$ directions to obtain the black hole in a D3-D3-P frame. Then, we rotate by an angle $\xi$ in the $(y_1,y_3)$ plane, and an angle $\omega$ in the $(y_2,y_4)$ plane. After these rotations, we T-dualize back along $y_3$ and $y_4$. Since these final T-dualities are not done strictly perpendicular or parallel to the stacks of D3-branes, the final D1-D5-P frame will also have D3 brane charges turned on in addition to the D1/D5/P charges. In the final D1-D5-P frame, we also add constants $b_{13}$ (resp. $b_{24}$) to the components $B_{13}$ (resp. $B_{24}$) of the NS 2-form; these components will also turn on additional D3 brane charges, which we will take to cancel the D3-brane charges generated by the T-dualities. More details on the whole dualization procedure can be found in appendix \ref{sec:appdualities}.

The final system is characterized by the six initial parameters $m, \delta_i, a_1, a_2$ as well as four new parameters: the angles $\xi, \omega$ and the constants $b_{13},b_{24}$. Our final D1-D5-P system is given by:
\begin{align}
\label{mod:metricwmod} ds^2 &= ds^2_6+ (H_1 H_2)^{1/2}\left[ H_{\xi}^{-1}\left(dy_1^2+dy_3^2\right) + H_{\omega}^{-1}\left(dy_2^2+dy_4^2\right)\right],\\
\label{mod:B2wmod} B_{NS} &= \left(b_{13} + H_{\xi}^{-1}(H_1-H_2)\sin\xi\cos\xi\right)dy_1\wedge dy_3\nn
 & + \left(b_{24} + H_{\omega}^{-1}(H_1-H_2)\sin\omega\cos\omega\right)dy_2\wedge dy_4,\\
 e^{2\phi} &= \frac{H_1 H_2}{H_{\xi}H_{\omega}},\\
\label{mod:C2wmod} C_2 &= \cos\xi\cos\omega K_1 + \sin\xi\sin\omega K_2,\\
 C_4 &= \left[H_2 K_2\sin\omega\cos\xi\ -H_1 K_1 \cos\omega\sin\xi \right]\wedge \frac{dy_1\wedge dy_3}{H_{\xi}}\nn
\label{mod:C4wmod}& + \left[H_2 K_2\cos\omega\sin\xi\ -H_1 K_1 \sin\omega\cos\xi \right]\wedge \frac{dy_2\wedge dy_4}{H_{\omega}},\\
\label{mod:C6wmod}C_6 &= -\left(H_2^2 K_2 \cos\xi\cos\omega+H_1^2 K_1\sin\xi\sin\omega\right)\wedge \frac{dy_1\wedge dy_2\wedge dy_3\wedge dy_4}{H_{\xi}H_{\omega}},
\end{align}
where we have used the new building blocks:
\begin{equation}
 H_{\xi} = H_1\sin^2\xi + H_2\cos^2\xi, \qquad H_{\omega} = H_1\sin^2\omega + H_2\cos^2\omega.
\end{equation}
Note that the position of the inner and outer horizons are obviously still given by (\ref{mod:horizonsinitial}).

As alluded to above, the system (\ref{mod:metricwmod})-(\ref{mod:C6wmod}) is not strictly a D1-D5-P system, but also contains two stacks of D3-branes (resp. wrapping $z,y_1,y_3$ and $z,y_2,y_4$). We will now calculate all of the charges present in the system (D1, D5, and two times D3). At this point it is important to realize that there are different notions of D-brane charge in type II supergravity due to the Chern-Simons terms in the action~\cite{Marolf:2000cb}. The physically relevant notion of charge is the Page charge, which is localized, conserved, and quantized. Page charge also transforms `covariantly' (with a slight abuse of language) under duality transformations~\cite{deBoer:2012ma}. In short, it is the D-brane Page charge which is most naturally connected with the charge of a stack of D-branes. We provide more details on Page charges in appendix \ref{app:page}, including the relevant formulas for computing them.

The Page D1,D5, and D3 charges of the system (\ref{mod:metricwmod})-(\ref{mod:C6wmod}) are given by:\footnote{One can wonder about picking up extra signs in these expressions due to our definition  (\ref{app:BSdef}) of Page currents, but these extra signs can be cancelled by picking an appropriate orientation of the manifolds at infinity over which we integrate to obtain the charges.}
\begin{align}
\label{mod:D5chargefull} Q_{D5}^P & = Q_2\cos\omega\cos\xi + Q_1\sin\omega\sin\xi,\\
Q_{D1}^P &= \cos\omega\cos\xi(Q_1 + b_{13} b_{24} Q_2) + \cos\omega\sin\xi(b_{13} Q_1 - b_{24} Q_2)\nn
\label{mod:D1chargefull} & + \sin\omega\sin\xi(Q_2+b_{13}b_{24} Q_1) + \sin\omega\cos\xi(b_{24}Q_1-b_{13} Q_2),\\
 Q_{D3(z13)}^P &= Q_2\left(b_{13}\cos\omega\cos\xi  - \cos\omega\sin\xi\right) \nn
\label{mod:D3chargefull1} & + Q_1\left( b_{13}\sin\omega\sin\xi + \sin\omega\cos\xi\right),\\
Q_{D3(z24)}^P &=  Q_2\left( b_{24}\cos\omega\cos\xi  - \sin\omega\cos\xi\right),\nn
\label{mod:D3chargefull2} & + Q_1\left(b_{24}\sin\omega\sin\xi+\cos\omega\sin\xi \right).
\end{align}
We wish for the background to be a 'pure' D1-D5-P system, that is, without D3-branes present. This is accomplished by demanding that (\ref{mod:D3chargefull1}) and (\ref{mod:D3chargefull2}) vanish. This gives us two constraints, fixing two of the parameters $(\xi,\omega,b_{13},b_{24})$:
\begin{align}
\label{mod:binfsol1} b_{13} &= \frac{Q_2\cos\omega\sin\xi-Q_1\sin\omega\cos\xi}{Q_2\cos\omega\cos\xi+Q_1\sin\omega\sin\xi} , \\
\label{mod:binfsol2} b_{24} &=\frac{Q_2\sin\omega\cos\xi-Q_1\cos\omega\sin\xi}{Q_2\cos\omega\cos\xi+Q_1\sin\omega\sin\xi}.
\end{align}
Even though we turn off the D3-charges in the D1-D5-P background black hole, we will still allow the probe tube to have D3-charges. In principle, the most general D1-D5-P multi-centered solution will contain arbitrary D3-charges on both of the centers. However, we are limiting ourselves to the 'pure' D1-D5 background which has zero D3-charges present. Note that we can always shift the D3-charge of the background in units of $Q_{D5}^P$ by performing a large gauge transformation of $B_{NS}$ (see appendix \ref{sec:apppagelargegauge}), but this will also have the effect of shifting the background Page D1-charge.

Finally, as the anti-self dual part of $B_{NS}$ gets fixed by the attractor mechanism in the near horizon region of the (near-)extremal geometry and is thus not a true moduli of the system~\cite{Larsen:1999uk}, we demand that we have self-dual $B_{NS}$:
\begin{equation}
\label{mod:selfdual} b_{13}=-b_{24},
\end{equation}
which immediately implies through (\ref{mod:binfsol1})-(\ref{mod:binfsol2}) that $\omega=-\xi$, so in the end we have one free parameter left to choose which determines how much of the moduli is turned on.

For a system where (\ref{mod:binfsol1})-(\ref{mod:binfsol2}) holds as well as (\ref{mod:selfdual}), we can further express the D1 and D5 Page charges (\ref{mod:D1chargefull}) and (\ref{mod:D5chargefull}) only in terms of the angle $\xi$ and the parameters $Q_1,Q_2$:
\begin{align}
\label{mod:D5chargewmod} Q_{D5}^P &= Q_2\cos\omega\cos\xi + Q_1\sin\omega\sin\xi = Q_2 \cos^2\xi - Q_1\sin^2\xi,\\
\label{mod:D1chargewmod} Q_{D1}^P &= \frac{Q_1 Q_2}{Q_2\cos\omega\cos\xi + Q_1\sin\omega\sin\xi} = \frac{Q_1 Q_2}{Q_2\cos^2\xi - Q_1\sin^2\xi}.
\end{align}
These expressions allow us to in principle find $Q_1$ and $Q_2$ for given charges $Q_{D1}^P,Q_{D5}^P$ and moduli angle $\xi$.

It will also be useful to have the expression for the moduli in terms of the Page D1/D5 charges $Q_{D1}^P,Q_{D5}^P$ and the angle $\xi$ (we choose the solution so that $b_{13}>0$):
\be \label{mod:bfuncpage} b_{13} = \csc2\xi\left(\sqrt{\frac{Q_{D1}^P}{Q_{D5}^P}\sin^2 2\xi+1}-\cos2\xi\right) .\ee
Finally, when we discuss the near-horizon decoupling limit, we will need the expression for the value of $B_{NS}$ at the horizon \emph{in the supersymmetric extremal limit} (note that of course $B_{24}^{(h-ext)}=-B_{13}^{(h-ext)}$), which can be read off from (\ref{mod:B2wmod}) using (\ref{mod:bfuncpage}):
\be \label{mod:bhorizonext} B^{(h-ext)}_{13} = \frac{Q_{D1}^P}{Q_{D5}^P}\sin2\xi \left(1+\frac{Q_{D1}^P}{Q_{D5}^P}\sin^2 2\xi\right)^{-1/2} .\ee

\subsection{Introducing the Probe Supertube}
It is in the background (\ref{mod:metricwmod})-(\ref{mod:C6wmod}) with (\ref{mod:binfsol1})-(\ref{mod:binfsol2}) and (\ref{mod:selfdual}) imposed, that we wish to introduce our probe (super)tube. In the D1-D5-P frame, the probe will have the usual KK-monopole dipole charge and D1/D5 monopole charges, but we will also turn on extra D3 monopole charges along the same directions as the D3-charges of the background (which are zero when (\ref{mod:binfsol1})-(\ref{mod:binfsol2}) holds). To calculate the action and Hamiltonian of this probe, we first dualize  our system (\ref{mod:metricwmod})-(\ref{mod:C6wmod}) to a F1-D2-D2 frame (where the D1's have become F1's). In this frame, the tube is a D4-brane dipole and can easily be studied using the standard DBI+WZ action for a D-brane probe. See appendix \ref{sec:appdualities} for a complete description of the dualities performed to obtain the final F1-D2-D2 frame.

The probe, being a dipole, will also wrap a non-compact direction $\alpha$, determined by two parameters $b_1,b_2$ as follows:
\begin{equation}
\label{mod:probecycle} \psi=b_1 \alpha, \qquad \phi = b_2\alpha.
\end{equation}
We will consider the probe at fixed $\theta$ for various radius $r$.

 Details of the calculation of the Hamiltonian of the tube (in the F1-D2-D2 frame) is given in appendix \ref{sec:appham}. Unfortunately, we did not obtain an analytic expression for the Hamiltonian, but can only study it numerically. We mention here that the tube Hamiltonian $\mathcal{H}$ is to be seen as a function of the radius $r$ at which it sits, its embedding constants $b_1,b_2$ defined above, its dipole charge $d$, its monopole charge D1-charge $q_1$, D5-charge $q_2$ and two D3 charges $q_3$ and $q_3'$:
\begin{equation}
\label{mod:endham}
 \mathcal{H} = \mathcal{H}(r, b_1, b_2,d,q_1,q_2,q_3,q_3').
\end{equation}
This Hamiltonian will be used in the following sections to study properties of the multi-centered D1-D5-P system with B-moduli turned on.


\section{Extremal Black Holes \& Supertubes} \label{Section:ExtremalBHs}
In this section, we will use the Hamiltonian (\ref{mod:endham}) to plot the interactions of a supertube probe and a supersymmetric BMPV black hole with the B-moduli turned on. We will always plot:
\be \label{extmod:tildeH} \tilde{\mathcal{H}}(\rho) = \mathcal{H}(\rho)-\mathcal{H}(\rho=0),\ee
where we have defined the new radial coordinate $\rho$ through:
\be \label{extmod:rho} \rho^2 = r^2 - r_+^2 .\ee
Of course, in the extremal case, we just have $\rho=r$. We will take the following parameters for the supertube embedding:
\be \label{extmod:embedding} b_1=1, \qquad b_2=0, \qquad \theta=0 .\ee

Thus, the important remaining parameters of our analysis will be the background (Page) charges $Q^P_{D1},Q^P_{D5},Q_3$, as well as the one anti self-dual angular momentum $J\equiv J_{\phi}=-J_{\psi}$, the probe (Page) charges $q_1,q_2,q_3,q_3'$, and the value of the B-moduli $b_{13}$. In particular, by specifying all of these parameters, the constants $Q_1,Q_2, \xi, \omega$ are all fixed as described in section \ref{sec:D1D5wmod}.

\subsection{The Same Probe at Different Moduli} \label{sec:extmodmoduli}
\begin{figure}[htbp]
\begin{center}
\includegraphics[width=0.6\textwidth]{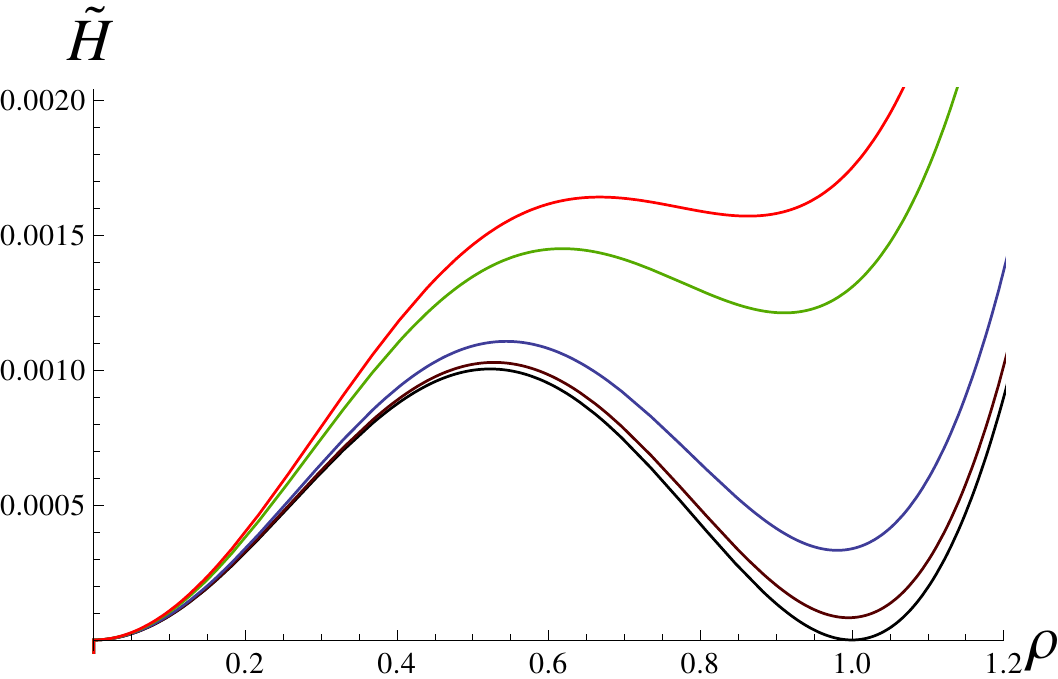}
\caption{The black hole supertube system with various values of the self-dual $B_{NS}$ moduli. The black hole charges are $Q_{D1}^P = 30, Q_{D5}^P = 20, Q_3=1, J=6$ while the supertube charges are $q_1= 2, q_2=1, d=1,q_3=q_3'=0$. The plots are for $b_{13} = 0,0.25,0.5,1,2$ (black, brown, blue, green, red, resp.).}\label{fig:extmodsimple}
\end{center}
\end{figure}

First of all, we can simply take a fixed probe with zero extra D3 charges (i.e. $q_3=q_3'=0$) and investigate its behavior at different values of the moduli parameter. We set $Q_{D1}^P = 30, Q_{D5}^P = 20, Q_3=1, J=6$ and $q_1= 2, q_2=1, d_3=1$, and plot for $b_{13} =0,.25,.5,1,2$ in figure \ref{fig:extmodsimple}.
At $b_{13}=0$ (the black graph), the moduli are turned off and we have the usual presence of the local supersymmetric minimum at:
\be \rho^2 = r_*^2 = q_1 q_2 - Q_3.\ee
However, for $b_{13}>0$, we see that the minimum lifts to a metastable value, implying in particular that there is no supersymmetric bound state of this tube and the black hole away from the zero moduli point. This is exactly as expected, since the D1-D5 black hole itself should be a true bound state~\cite{Dhar:1999ax}; the D1-D5 system at finite moduli is not marginally bound.

\begin{figure}
\begin{center}
\includegraphics[width=0.6\textwidth]{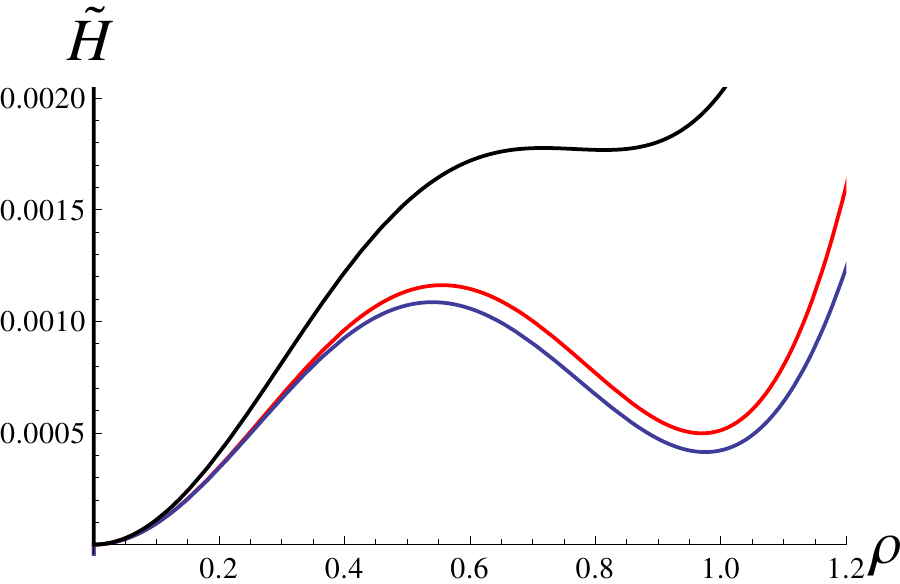}
\caption{The black hole supertube system with various values of the self-dual $B_{NS}$ moduli. The black hole and supertube charges are the same as in fig. \ref{fig:extmodsimple}. The three plots are for $b_{13} =0.61$ (corresponding to $\xi=\pi/12$; red); $b_{13}=4.08$ ($\xi=y=\pi/2-\pi/12$; blue); and $b_{13}=1.58$ ($\xi=\pi/4$; black).}
\label{fig:extmodmaxlifting}
\end{center}
\end{figure}

However, if we increase the moduli parameter $b_{13}$ further and further for the same probe, we notice that the lifting of the minimum starts going \emph{down} again. In fact, the value of $b_{13}$ at which the lifting is maximal corresponds (see (\ref{mod:bfuncpage})) to the angle $\xi=\pi/4$; see fig. \ref{fig:extmodmaxlifting}.

We can explain this as follows. For an extremal D1/D5-system, the near-horizon decoupling geometry is the well-known $AdS_3$ throat~\cite{Cvetic:1998xh}; in appendix \ref{sec:apphamdecouple} we give some details on the decoupling limit. In this near-horizon geometry, the value of the B-moduli is given by (\ref{mod:bhorizonext}), which is the B-moduli parameter in the dual D1/D5 CFT~\cite{Larsen:1999uk}. This expression (\ref{mod:bhorizonext}) is manifestly invariant under $\xi\rightarrow \pi/2-\xi$ and reaches its maximum at $\xi=\pi/4$. Thus, even though we can crank up $b_{13}$ to infinity (corresponding to $\xi\rightarrow\pi/2$ in (\ref{mod:bfuncpage})), the value of the near-horizon B-moduli remains bounded by its value at $\xi=\pi/4$ (which is also manifest from (\ref{mod:bhorizonext})), as appears to be the case in fig. \ref{fig:extmodmaxlifting}. We propose that the lifting of the minimum is controlled by the value of the near-horizon B-moduli (so is maximal when $\xi=\pi/4$); further evidence for this is provided by the fact that the decoupling limit Hamiltonian $\mathscr{H}$ has exactly the symmetry $\xi\rightarrow \pi/2-\xi$. In the appendix \ref{sec:apphamdecouple} we explain the calculation of the decoupling limit Hamiltonian $\mathscr{H}$ and how we verified this symmetry explicitly.

It is interesting to note that the maximum value of the near-horizon B-moduli that we can obtain (i.e. with $\xi=\pi/4$) is:
\be (B^{(h-ext,max)}_{13})^2 = \left( \frac{Q_{D1}^P}{Q_{D5}^P} \right)^2 \left(1+\frac{Q_{D1}^P}{Q_{D5}^P}\right)^{-1},\ee
which is always strictly smaller than the maximum value allowed in principle for $B^2$ in the near-horizon geometry, namely $Q_{D1}^P/Q_{D5}^P$~\cite{Larsen:1999uk, Dijkgraaf:1998gf}. This means there is a window of allowed values of $B^{(h-ext,max)}_{13}$ that we cannot explore using our methods.

The reason why we cannot probe the full range of allowed $B^{(h-ext,max)}_{13}$ values can be understood from the duality procedure that we use to turn on the moduli in the background. To give an intuitive way of understanding this: we start with only two stacks of branes (in the initial D1-D5 frame), which are then rotated and dualized etc. as explained in the text. The resulting D3-brane Page charges in the final D1-D5 system are given by (\ref{mod:D3chargefull1}) and (\ref{mod:D3chargefull2}) are thus obviously related to the initial stacks; in particular, these charges are not arbitrary or unlimited in the final D1-D5 frame. Since the resulting moduli parameters are then tuned (by introducing the constants $b_{13},b_{24}$) such that the D3-charges (\ref{mod:D3chargefull1})-(\ref{mod:D3chargefull2}) are zero, this implies that the moduli parameters are themselves also not arbitrarily chosen and are limited by this procedure. One might think that introducing extra stacks of initial branes might solve the problem, but such solutions with extra stacks of D-branes that would affect the resulting B-moduli are not known or non-existent in general. In any case, even though we cannot explicitly study systems using our techniques with B-moduli larger than the bound given above, we do not expect any phenomena that we study in this paper to be affected qualitatively by increasing the B-moduli beyond our bound, as all phenomena that we study are ``smooth'' in the B-moduli parameter and give no indication of changing dramatically at any value of the B-moduli. 

Finally, we also note that in the graph in figure \ref{fig:extmodmaxlifting} of the Hamiltonian, plotting a tube at a certain value of $\xi$ and $\pi/2-\xi$ reveals that, even though the lifting of the minimum is more or less the same for the two systems, the graphs do not completely coincide. This we can ascribe to the effects of the different asymptotically flat system that the two tubes are embedded in. Even though the near-horizon geometry of both of these asymptotically flat geometries coincides, they are different flat geometries; this affects the tube differently at large $r$. Using loose holographic language, the two different asymptotically flat geometries with $\xi$ and $\pi/2-\xi$ correspond to turning on two different irrelevant operators in the same dual CFT state.


\subsection{Finding a Supersymmetric Bound Probe at Finite Moduli (Or Not)}

The next obvious question is whether for a given D1-D5-P black hole background at a given moduli, there is a probe tube which is supersymmetric w.r.t. this background. We can view the background charges and moduli as fixed, together with the probe D1 and D5 charges, and ask whether there exists values for the probe D3 charges that makes the probe mutually supersymmetric with the background. Indeed, the D3-charges $q_3,q_3'$ of the supersymmetric probe are given by the $Q^P_{D3}$ in (\ref{mod:D3chargefull1}) and (\ref{mod:D3chargefull2}), where $Q_1$ and $Q_2$ are the solutions to the equations (\ref{mod:D5chargewmod}) and (\ref{mod:D1chargewmod}) with $Q_{D1}^P=q_1$ and $Q_{D5}^P=q_2$ (i.e. the probe Page D1/D5 charges). Choosing the probe D3-charges in this way assures us that the probe supersymmetry projectors are compatible with those of the background (see e.g. \cite{Dabholkar:2009dq}). The result, as seen in fig. 4, is a probe with a supersymmetric minimum.

\begin{figure}[htbp]
\begin{center}
\includegraphics[width=0.5\textwidth]{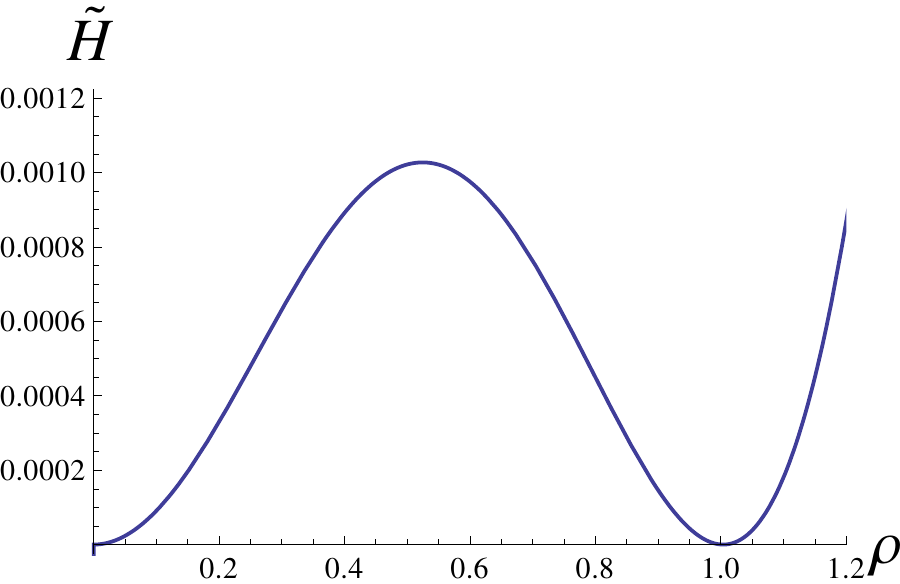}\\
 \small{$q_1=2,q_2=1,q_3=-q_3'=0.09$}\\
 \small{$Q_{D1}^P = 30, Q_{D5}^P = 20, b_{13}=0.5$, $Q_3=1, J=6$ }\\
\caption{We see that the probe with these particular charges has a supersymmetric minimum. However, for a given background and given $q_1,q_2$, the D3-charge that the supersymmetric probe has is completely fixed and is in general, like here, not quantized.}
\label{fig:sugraModifiedTube}
\end{center}
\end{figure}
However, we must remember that we expect the Page charge to be a \emph{quantized} number in the full quantum theory. For generic (quantized) fixed tube charges $(q_1,q_2)$, the D3-charges $q_3,q_3'$ that are needed to obtain a supersymmetric minimum as in fig. \ref{fig:sugraModifiedTube} will not be an integer, and so is not allowed in the full quantum theory. This means that, in general, there is no possible allowed tube probe with given D1 and D5 charges $(q_1,q_2)$ that will have a supersymmetric bound state with the background black hole; turning on the $B_{NS}$ moduli makes the system become metastable and therefore non-supersymmetric.

This mechanism of supersymmetry breaking due to non-quantization of the charges is completely analogous to the weak-coupling F1-P picture discussed in the introduction of this paper. Note that the role of the tilting of the torus in the F1-P discussion is played by the B-field in our D1-D5 + probe system.

\subsection{Qualitative Amount of Breaking} \label{sec:extmodqual}
We would like to get a qualitative feeling of how `much' supersymmetry is broken for the BH-tube system when the B-moduli is turned on. We will investigate two scalings to further our understanding of this. In this subsection, we always have zero D3-charges on the probe, $q_3=q_3'=0$.

First of all, we discuss how to qualitatively discuss how `much' supersymmetry is broken. We propose to compare two parameters: the height of the minimum of the potential at finite $r$, and the height of the potential barrier (the `hump' separating the horizon and this minimum). For example, in fig. \ref{fig:sugraModifiedTube}, the minimum is at 0 since it is a supersymmetric minimum, while the maximum is around 0.001. The ratio of these two parameters quantifies the stability of the minimum in the potential (i.e. tunneling probabilities will depend on this ratio\footnote{One could argue that a better parameter instead of the height of the `hump' which quantifies the tunneling probability would be the area under the potential curve between the minimum and the horizon; using this quantity would give the same conclusions for all systems that we consider.}). For a supersymmetric probe, the ratio of these two parameters will always be zero; for a non-supersymmetric probe, we will use the departure of this ratio from zero to give us a measure for the lifting of the marginal state, i.e. to quantify how `much' supersymmetry is broken.

\begin{figure}
\begin{center}
\begin{tabular}{cc}
\includegraphics[width=0.48\textwidth]{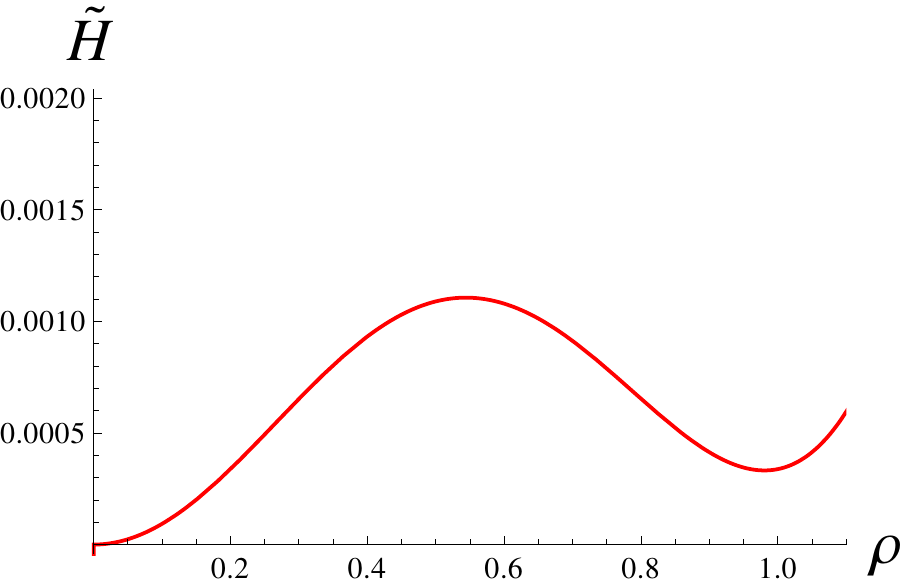} & \includegraphics[width=0.48\textwidth]{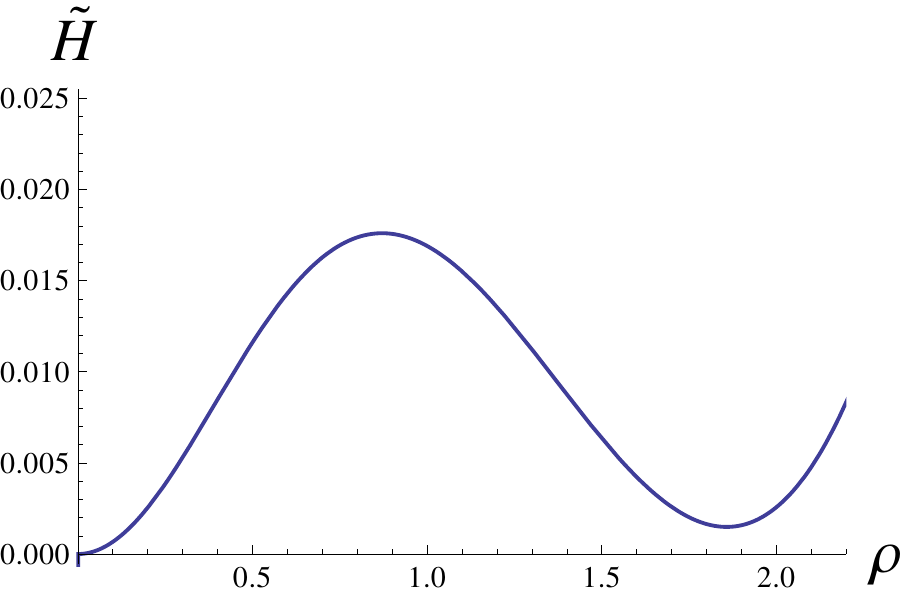}\\
(a) $\alpha = 1$ & (b) $\alpha=1.5$\\
 \includegraphics[width=0.48\textwidth]{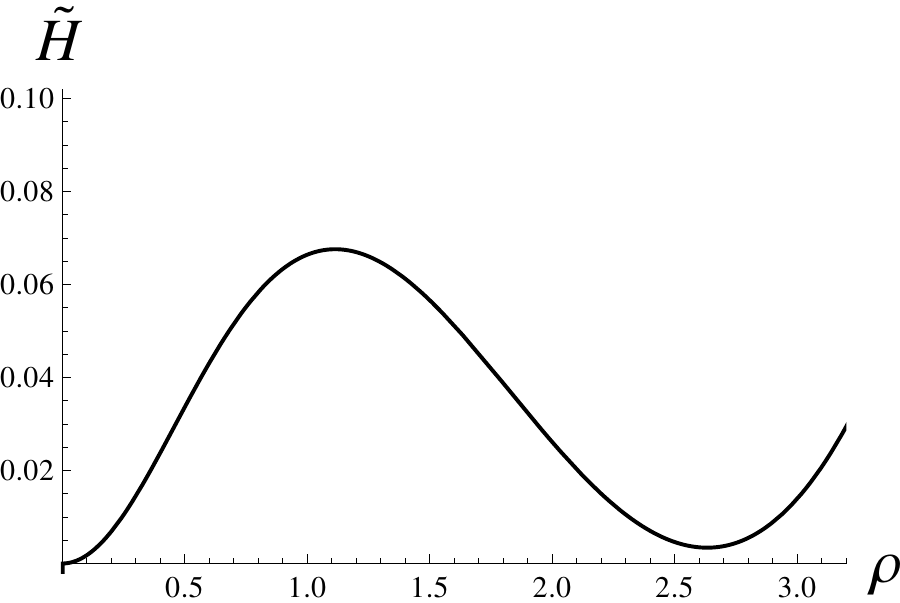} &  \includegraphics[width=0.48\textwidth]{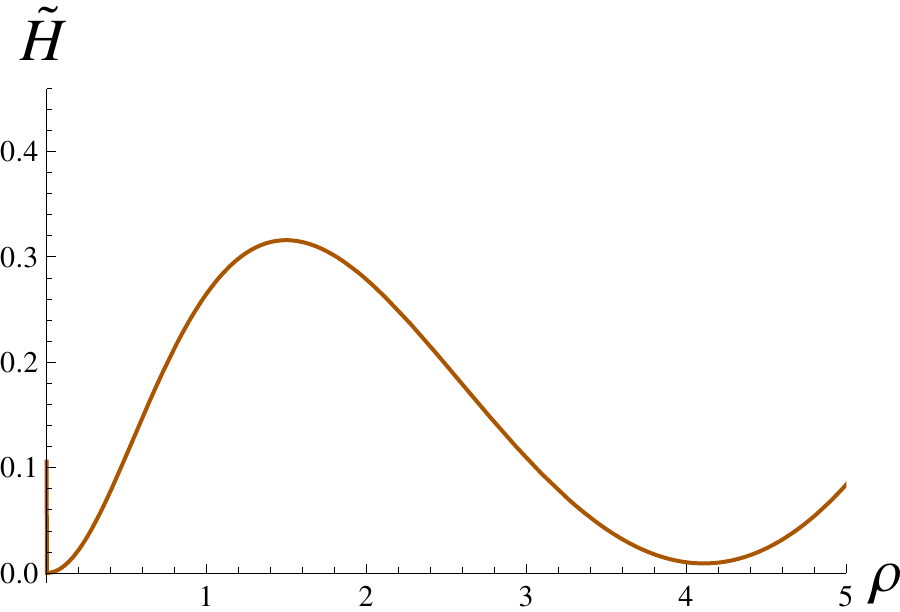}\\
(c) $\alpha=2$ & (d) $\alpha=3$
\end{tabular}
\caption{The black hole background charges are the same as in fig. \ref{fig:extmodsimple}; for the moduli we take $b_{13}=0.5$. The tube probe charges are $q_1= 2\alpha; q_2 = \alpha$ (for $\alpha=1; 1.5; 2; 3$).}
\label{fig:extmodalphascaling}
\end{center}
\end{figure}

An interesting question to ask is what the effect would be when we have charges on the probe of the same order as on the background. While this is per definition outside of the probe limit, we can still investigate this effect somewhat by keeping the background charges fixed and increasing the probe charges to see if any trend becomes clear. We will use the following scaling for the probe charges:
\be q_1 \rightarrow \alpha\, q_1, \qquad q_2\rightarrow \alpha\, q_2.\ee
A trend for this scaling is clearly visible in figure \ref{fig:extmodalphascaling}: the supersymmetry breaking effect goes down with increasing supertube charges. In the limit where both centers have large and comparable charges (large in the sense that we can approximate the two-centered solution by its backreacted configuration in supergravity), we expect the lifting of the marginal state to a metastable state to become negligible.

Note that the radius at which the minimum of the supertube for large probe charges ``settles down'' seems to be the supersymmetric radius, $r_*=\sqrt{q_1 q_2 - Q_3}$. We also note that even though our natural parameter for quantifying the supersymmetry breaking (i.e. the ratio of the height of the minimum of the potential to the height of the potential barrier maximum) decreases as the charges increase with $\alpha$, at the same time the absolute value of the minimum of the potential \emph{does not} decrease and in fact increases. Clearly, only a fully backreacted SUGRA two-centered solution will give us the complete and unambiguous picture of the fate of these two centers when they have large and comparable charges.

We can also consider the scaling of all quantities with the string coupling constant. The charges and radius will scale as:
\begin{align}
Q_{D1}^P &\rightarrow \eta\, Q_{D1}^P,& Q_{D5}^P &\rightarrow \eta\, Q_{D5}^P,\\
Q_3 &\rightarrow \eta^2\, Q_3, & J&\rightarrow \eta^2\, J,\\
q_1 & \rightarrow \eta\, q_1, & q_2 &\rightarrow \eta\, q_2,\\
r & \rightarrow \eta\, r. &
\end{align}
Note that the first line implies that the parameters $Q_{1,2}\rightarrow \eta\,  Q_{1,2}$ as well. The value of $B^{h-ext}$ and the angle $\xi$ are unchanged (for fixed $b_{13}$) under this scaling.

\begin{figure}[h]
\begin{center}
\begin{tabular}{cc}
\includegraphics[width=0.5\textwidth]{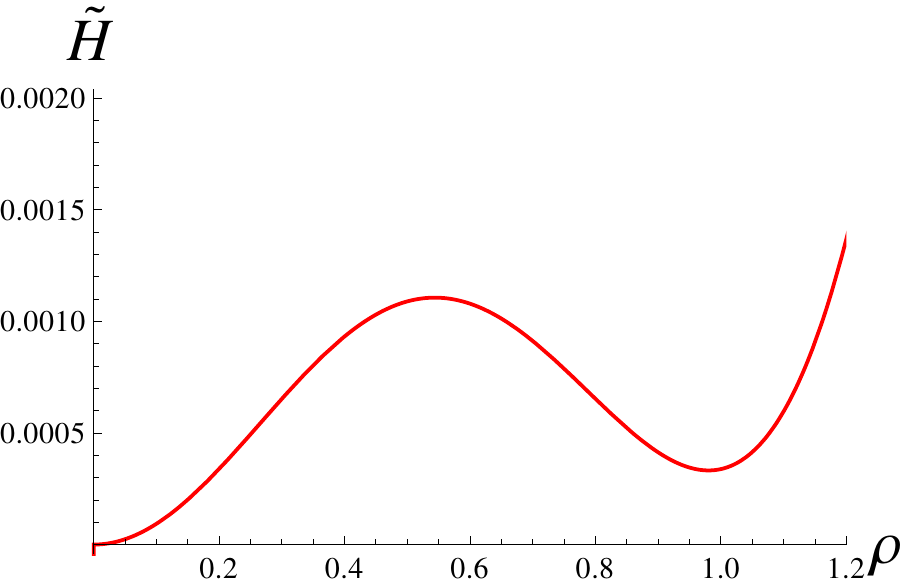} & \includegraphics[width=0.5\textwidth]{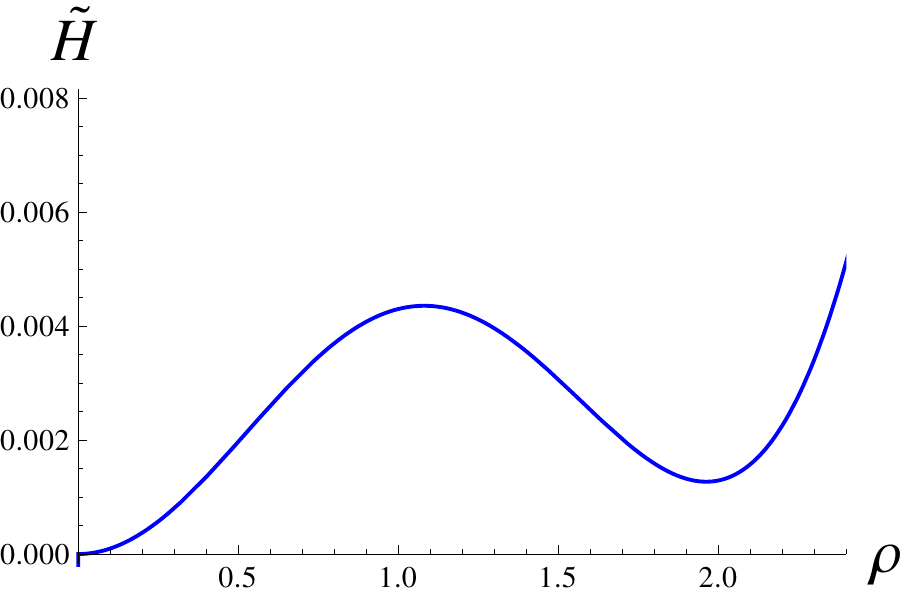}\\
(a) $\eta = 1$ & $\eta = 2$
\end{tabular}
\caption{The black hole charges are $Q_{D1}^P =  30\, \eta, Q_{D5}^P =  20\, \eta, Q_3=1\,\eta^2, J=6\, \eta^2$ while the tube charges are $q_1= 2\, \eta , q_2=1\,\eta , d_3=1$. The moduli parameter is set to $b_{13}=0.5$.}
\label{fig:extmodetascaling}
\end{center}
\end{figure}

In the decoupling limit at zero moduli, the Hamiltonian scales as $\eta^2$ with this scaling~\cite{Chowdhury:2011qu}:
 \be \mathscr{H}(\eta\, Q_{D1}^P, \eta\, Q_{D5}^P, \eta^2\, Q_3, \eta^2\, J, \eta\, q_1, \eta\, q_2, \eta\, r) = \eta^2\, \mathscr{H}(Q_{D1}^P,  Q_{D5}^P,  Q_3,  J, q_1,  q_2, r).\ee
Figure \ref{fig:extmodetascaling} reveals that the finite moduli decoupling limit Hamiltonian should also enjoy this scaling. Indeed, we were able to prove this scaling analytically in appendix \ref{sec:apphamdecouple}. 

Thus, even though the absolute value of the potential scales with $g_s$, it is clear that the amount of supersymmetry breaking (as measured by the ratio of the height of the minimum to the height of the maximum of the potential) does not depend on $g_s$. This gives a strong indication that an analogous analysis of the lifting of states as we are doing could be performed in the weak coupling CFT picture with similar results.\footnote{See Ref~\cite{Burrington:2012yq} for recent results on perturbing the weakly coupled CFT which demonstrate lowering of anomalous dimensions of some stringy states.}

\section{Non-extremal Black Holes \& Tube Instabilities} \label{Section:NonExtremalBHs}
In this section, we will again plot the shifted Hamiltonian $\tilde{\mathcal{H}}(\rho)$ given by (\ref{extmod:tildeH}) with $\rho$ given by (\ref{extmod:rho}), for a few supertube configurations in \emph{non-}extremal background black hole setups. The embedding of the tube will still be given by (\ref{extmod:embedding}). The black hole charges that we need to specify are the Page charges $Q^P_{D1},Q^P_{D5},Q_3$, the angular momenta $J_{\psi},J_{\psi}$ and the non-extremality parameter $m$. The probe (Page) charges to specify are still $q_1,q_2,q_3,q_3'$. Finally, we also need to specify the B-moduli $b_{13}$. Specifying all these parameters fixes the constants $\delta_i,a_j,\xi,\omega$ as well as the horizon radius $r_+$, as described in section \ref{sec:D1D5wmod}.

In~\cite{Chowdhury:2011qu}, the system of non-extremal D1-D5-P black holes with supertube probes was analyzed at zero moduli. The existence of truly bound states (i.e. where the energy of the system where the supertube is at some finite radius $r_*>r_+$ is lower than the energy of the system with the supertube at the horizon) was discovered, leading to the claim that certain non-extremal black holes have an instability towards ``spitting out'' supertubes and thus lowering their energy.

A relevant question is whether these bound states found in~\cite{Chowdhury:2011qu} at zero moduli will persist at finite B-moduli. The answer can be given straightforwardly: some will, some will not.

\begin{figure}[htbp]
\begin{center}
\includegraphics[width=0.6\textwidth]{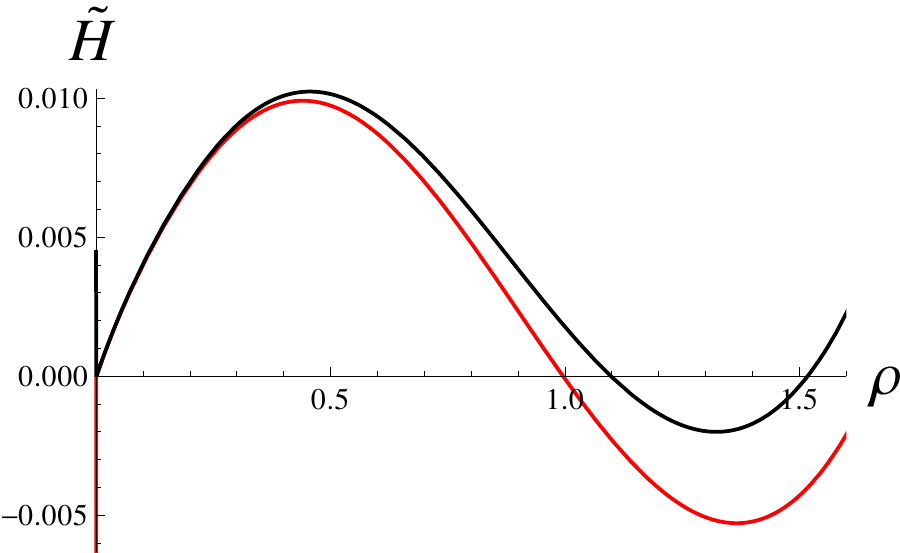}
\caption{The non-extremal black hole system with charges $Q_{D1}^P = 30, Q_{D5}^P = 20, Q_3=0.1, J_{\psi}=3,J_{\phi}=0, m=0.1$ and supertube charges $q_1= 2, q_2=1, d=1,q_3=q_3'=0$. The plots are for the moduli values $b_{13} = 0$ (red) and $b_{13}=1.58$ (black), corresponding to $\xi=0$ and $\xi=\pi/4$, respectively.}
\label{fig:nonextmodnotlifted}
\end{center}
\end{figure}

In figure \ref{fig:nonextmodnotlifted} we give one example of a bound state which remains bound at the maximum value of the near-horizon moduli $B^{(h)}_{13}$ that we can turn on, which corresponds to $\xi=\pi/4$ and is the value at which we expect the maximum amount of lifting (see section \ref{sec:extmodmoduli}). As is obvious from fig. \ref{fig:nonextmodnotlifted}, the binding energy of the bound state decreases when the moduli are turned on, but does not shrink to zero.

\begin{figure}[htbp]
\begin{center}
\includegraphics[width=0.6\textwidth]{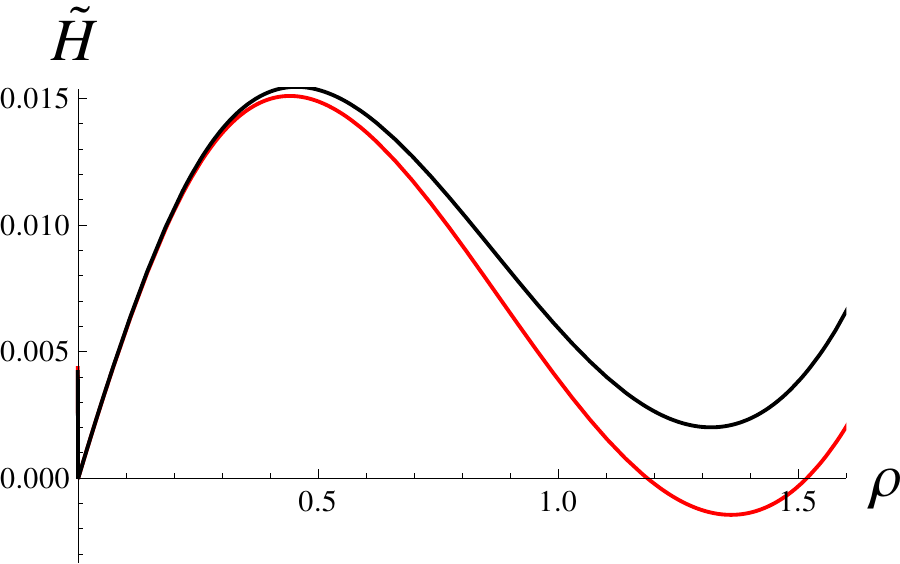}
\caption{The non-extremal black hole system with all the same black hole and supertube charges as in fig. \ref{fig:nonextmodnotlifted} except that the angular momenta here are  $J_{\psi}=5,J_{\phi}=-5$. The plots are for the moduli values $b_{13} = 0$ (red) and $b_{13}=1.58$ (black), corresponding to $\xi=0$ and $\xi=\pi/4$, respectively.}
\label{fig:nonextmodlifted}
\end{center}
\end{figure}

On the other hand, in figure \ref{fig:nonextmodlifted} we give an example of a bound state at zero moduli which lifts to a meta-stable state at finite moduli.

In general, the trend is clear from fig. \ref{fig:nonextmodnotlifted} and fig. \ref{fig:nonextmodlifted}: the binding energy of the bound state will decrease, but whether it shrinks entirely to zero or not depends on the bound state in question. The take-home message is this: the entire analysis of~\cite{Chowdhury:2011qu} can be repeated at finite moduli with similar results, and we expect non-extremal black hole states which are unstable towards ``spitting out'' supertubes to exist at all values of the B-moduli.

One important caveat to the above discussion is the fact that we are unable to scan the full range of B-moduli, but rather are limited to a maximum near-horizon value $(B^{(h,max)}_{13})$ (see section \ref{sec:extmodmoduli})
which is always strictly smaller than the maximum value allowed in principle in the theory~\cite{Dijkgraaf:1998gf}. However, based on the above analysis, we don't expect this ``extra'' range of B to affect our conclusion that such non-extremal BH-supertube bound states exist at every value of the B-moduli.

\section{Conclusion} \label{Section:Conclusion}
We have initiated a study of the fate of multi-centered solutions away from the special surface in the moduli space where the constituent branes can split with no energy cost~\cite{Seiberg:1999xz}.  More precisely, we studied the bound state of a (non-)extremal D1-D5 black hole and a probe supertube after turning on a self-dual $B_{NS}$ field on the torus that the D5-branes wrap using methods studied in~\cite{Maldacena:1999mh,Dhar:1999ax,Lunin:2003gw}.

We were motivated in studying this problem because of the phenomenon of entropy enigma in the D1-D5 system --  certain supersymmetric multi-centered solutions were found to have more entropy than the elliptic genus~\cite{Bena:2011zw,Chowdhury:2012ff} -- and a study of entropy enigma for $\mathcal N=4$ supergravity~\cite{Dabholkar:2009dq} that  suggested that such enigmatic solutions would become non-supersymmetric away from special submanifolds in the moduli space.  In section \ref{Section:ExtremalBHs} we realized this breaking of supersymmetry by observing that the bound state of the two centers became metastable and thus non-supersymmetric.\footnote{This supersymmetry breaking essentially arose because of the charges being quantized and by artificially making the charges quantized we could make the bound state marginally bound again.}
While this confirms that the enigmatic solutions are not captured by the elliptic genus as they are not supersymmetric everywhere in the moduli space \cite{Dabholkar:2009dq}, one may wonder what the fate of fully backreacted multi-centered bound states (of e.g.~\cite{Bena:2011zw,Chowdhury:2012ff}) would be when generic moduli are turned on. The common lore is that the mass of non-supersymmetric states is not protected and at strong coupling they lift. However, we were interested in asking how much they lift and if such multi-centered solutions appear discontinuously in the moduli space. By making the mass of the probe comparable to the background, it appears that we can make the lifting become negligible. While this by definition takes us out of the probe approximation, it gives strong indication that the fully backreacted states of supergravity will be stable against change of moduli which make them non-supersymmetric.

In~\cite{Chowdhury:2011qu} a Penrose process like instability of non-extremal rotating black holes towards spitting out supertubes was found. One may wonder if such an instability is an artifact of the branes being marginally bound in the absence of the self-dual $B_{NS}$ moduli.\footnote{We thank Shiraz Minwalla for bringing this possibility to our attention.} In section \ref{Section:NonExtremalBHs} we repeated the analysis of~\cite{Chowdhury:2011qu} after turning on the moduli and found that while the instability is decreased in line with the intuition of the branes being attracted at non-zero moduli, it does still persist at non-zero moduli.

A caveat in our analysis is that the maximum value of the self dual $B_{NS}$ field that we can turn on using the procedure of~\cite{Maldacena:1999mh,Dhar:1999ax,Lunin:2003gw} is less than the maximum value allowed by the D1-D5 CFT. The source of this deficit is explained in a footnote in section \ref{sec:extmodmoduli}. It would be interesting to figure out a way to turn on the full range of the self dual $B_{NS}$ but we do not expect it to change our conclusions as all the phenomena we have studied are ``smooth'' in the B-modulus and give no indication of changing dramatically when the modulus is increased further.

We would like to comment on the implications of our results for the fuzzball proposal~\cite{Mathur:2005zp,Bena:2004de,Skenderis:2008qn,Balasubramanian:2008da,Chowdhury:2010ct}. The strong form of the proposal~\cite{Bena:2007kg} is that
\begin{quote}
\dots among the typical black hole microstates there are smooth solutions that can be described using supergravity.
\end{quote}
and has led to a rich literature trying to find most, if not all, of the microstates accounting for the Bekenstein-Hawking entropy of a black hole by constructing multi-centered microstates (see~\cite{Bena:2007kg} for a review and~\cite{Bena:2008nh,Bena:2010gg,Bena:2011uw,Bena:2011fc,Bena:2012zi} for more recent work.). A property of the supersymmetric BMPV black hole is that its entropy is equal to the elliptic genus and this means that its microstates should remain supersymmetric under change of moduli.\footnote{We would like to thank Atish Dabholkar for clarifying this issue to us.} We have found that two-centered configurations where one is a black hole become non-supersymmetric at finite B-moduli. It would be interesting to study the fate of the various multi-centered smooth configurations and to classify which of them remain supersymmetric under change of moduli. While some of them may be amenable to probe analysis, many would require the generalization of the techniques of this paper to the fully backreacted case. Generalizing these techniques is at first glance rather difficult as we are unable to identify e.g. the combinations of charges that appear in the Hamiltonian (e.g. symplectic products) - if such ``clean'', invariant combinations will even exist in this case -  as our calculation of the Hamiltonian was purely numerical. We hope to return to this issue in the future.

Another direction of interest would be to do a parallel analysis of multi-centered configurations at the orbifold point. At the orbifold point the dual to the configurations we studied here involves a division of the total winding of the D1-D5 CFT into a long string sector and a short string sector~\cite{Bena:2011zw,Chowdhury:2012ff}. In this paper we found evidence that when the two centers are comparable (in their masses and charges for instance) then the lifting of states from supersymmetry breaking coming from non-zero moduli becomes negligible. The dual to the self-dual B-field considered here are known to be marginal deformations involving $\mathbb Z_2$ twist operators at the orbifold point~\cite{David:2002wn}. It would be interesting to see if the lifting of states persists in a similar manner at the orbifold point and if the lifting goes down with increasing the winding in the long string sector using techniques developed in~\cite{Gava:2002xb,Avery:2010er,Avery:2010hs,Avery:2010vk,Burrington:2012yq}. Indeed, a first strong indication that similar features should be found at the weak-coupling orbifold point is given by the analysis we performed in section \ref{sec:extmodqual}, which suggests that the lifting of states is independent of $g_s$.

Finally, we would like to comment that we would expect similar breaking of supersymmetry on turning on certain moduli in the four dimensional $\mathcal N=2$ multi-center solutions. Indeed, this is the setting in which entropy enigma was discussed initially in~\cite{Denef:2007vg} and the enigmatic states not contributing to the elliptic genus was discussed in~\cite{Dabholkar:2009dq}. Some of these configurations can be viewed as dimensional reduction of the D1-D5 configurations and so we definitely expect similar physics. However, in that case there are many more moduli and it is not clear how and which of them to turn on to see the physics of SUSY breaking. We hope to come back to this interesting issue in the future.

\section*{Acknowledgments}
We would like to thank M. Baggio, I. Bena,  J. de Boer, C. Closset, A. Dabholkar, F. Larsen, S. Mathur, S. Minwalla, S. Murthy, A. Puhm, B. Sahoo, M. Shigemori and B. Vercnocke,   for stimulating discussions. The work of
BDC is supported by the ERC Advanced Grant 268088-EMERGRAV. This work is part of the research programme of the Foundation for Fundamental Research on Matter (FOM), which is part of the Netherlands Organisation for Scientific Research (NWO).

\appendix
\section{Conventions}\label{app:conventions}
In this appendix we will spell out the various conventions that we use.
\subsection{IIA/IIB Supergravities}
We use IIB supergravity to talk about the 5D black hole in the D1/D5 frame, and we use IIA supergravity when we are calculating the D4-brane dipole probe Hamiltonian in the final F1-D2-D2 frame. Thus, it is useful to mention our conventions for both supergravities.

\subsubsection{IIA Supergravity}
Our conventions for the IIA action are:
\begin{align}
S_{IIA} &= \frac{1}{2\kappa_0^2} \int d^{10}x \sqrt{-g}\left\{ e^{-2\Phi}\left[R + 4(\nabla\phi)^2 - \frac{1}{12} H_3^2\right]\right.\nn
&\left. -\frac14 F_2^2 - \frac{1}{48}F_4^2\right\} - \frac{1}{4\kappa_0^2}\int B_2\wedge dC_3\wedge dC_3\nn
&= \frac{1}{2\kappa_0^2} \int \left\{ e^{-2\Phi}\left[ -*R -4* d\Phi\wedge d\Phi + \frac12 *H_3\wedge H_3\right]\right.\nn
& \left. +\frac12 *F_2\wedge F_2 + \frac12 *F_4\wedge F_4\right\} - \frac{1}{4\kappa_0^2} \int B_2\wedge dC_3\wedge dC_3.
\end{align}
Here we use the definition for the RR form-fields and potentials:
\be F_p = dC_{p-1} + H_3\wedge C_{p-3},\ee
and we define the Hodge duals:
\begin{align}
\label{app:IIAduality1} F_6 &= -*F_4, & F_4 &= *F_6,\\
 \label{app:IIAduality2} F_8 &= + *F_2, & F_2 &= -*F_8,
\end{align}
since then the equations of motion and Bianchi identities for $C_p$ can be summarized by:
\be \label{app:IIAEOM} dF_p + H_3\wedge F_{p-2}  = 0. \ee
The equation of motion and Bianchi identity for $B_2$ are:
\begin{align} d(e^{-2\Phi}*H_3) -F_2\wedge F_6 +\frac12 F_4\wedge F_4 &=0 ,\\
 dH_3 &=0.
\end{align}

We can also derive the equations of motion and Bianchi identities from the following so-called democratic formulation of IIA supergravity, which treats all form fields on equal footing:
\begin{align}
S_{IIA,dem} =\frac{1}{2\kappa_0^2} \int &\left\{ e^{-2\Phi}\left[ -*R -4* d\Phi\wedge d\Phi + \frac12 *H_3\wedge H_3\right]\right.\nn
\label{appconv:IIAdemocratic} & \left. +\frac14 *F_2\wedge F_2 + \frac14 *F_4\wedge F_4 + \frac14 *F_6\wedge F_6 + \frac14 *F_8\wedge F_8\right\}.
\end{align}
Note the absence of Chern-Simons terms in this formulation, as well as the presence of an extra factor of $1/2$ that is present because we are in a sense `double counting' all terms in the action. This action gives us the equations of motion (\ref{app:IIAEOM}), but we need to impose the duality conditions (\ref{app:IIAduality1})-(\ref{app:IIAduality2}) by hand in addition.

For a (single) D-brane source, the worldvolume action is:
\begin{align}
\label{appconv:Dbrane} S_{Dp} &= S_{DBI} + S_{WZ},\\
S_{DBI} &= -T_{Dp}\int d^{p+1}\xi e^{-\phi}\sqrt{-\det\left(g+B+F\right)},\\
S_{WZ} &= T_{Dp} \int \left( C_{p+1} + C_{p-1}\wedge(B+F) + \frac12 C_{p-4}\wedge(B+F)\wedge(B+F) + \cdots \right)\nn
& = T_{Dp} \int e^{B+F} C_{p+1}.
\end{align}
Note that $F=2\pi\alpha' \mathcal{F}$, where $\mathcal{F}=d\mathcal{A}$ is the usual definition of the gauge field on the brane. The quantities $g, B, C_p$ in this action should all be viewed as pulled-back on the D-brane worldvolume.

\subsubsection{IIB Supergravity}
Our action is given by:
\begin{align} S_{IIB} &= \frac{1}{2\kappa_0^2} \int d^{10}x \sqrt{-g}\left\{ e^{-2\Phi}\left[R + 4(\nabla\phi)^2 - \frac{1}{12} H_3^2\right]\right.\nn
&\left. -\frac12 F_1^2-\frac{1}{12} F_3^2 - \frac{1}{480}F_5^2\right\} + \frac{1}{4\kappa_0^2}\int C_4\wedge dC_2\wedge H_3.
\end{align}
Again, the RR form-fields and potentials are given by:
\be F_p = dC_{p-1} + H_3\wedge C_{p-3}.\ee
The duality relations for IIB are:
\begin{align}
 F_5&=*F_5,&&\\
 F_3 &= -*F_7, & F_7 &= -*F_3,\\
 F_1 &= +*F_9, & F_9 &= +*F_1.
\end{align}
The equations of motion are then again given by (\ref{app:IIAEOM}), so:
\be dF_p + H_3\wedge F_{p-2}  = 0. \ee
The equation of motion and Bianchi identity for $B_2$ are:
\begin{align} d(e^{-2\Phi}*H_3) + F_1\wedge F_7 + F_5\wedge F_3 &=0 ,\\
 dH_3 &=0.
\end{align}

The IIB action can also be given in the democratic formulation, the action for the RR fields is just:
\be 2\kappa_0^2 S_{IIB(RR),dem} = +\frac14\left(*F_1\wedge F_1 + *F_3\wedge F_3 + *F_5\wedge F_5 +  *F_7\wedge F_7 + *F_9\wedge F_9\right). \ee

\subsection{Duality Rules}
Here we list the T-duality and S-duality rules and conventions that we use for reference and convenience.

\subsubsection{T-duality}
In our conventions for IIA/B supergravity as detailed above, a T-duality along the $y$ direction is given by the following formulas for the NS fields~\cite{Johnson:2000ch}, where $\mu,\nu,\rho,\sigma, \cdots$ always run over the spacetime indices along which we are not dualizing:
\begin{align}
g_{\mu\nu}' & = g_{\mu\nu} - \frac{g_{\mu y} g_{\nu y}-B_{\mu y} B_{\nu y}}{g_{yy}},\\
g_{\mu y}' & = \frac{B_{\mu y}}{g_{yy}},\quad  g_{yy}' = \frac{1}{g_{yy}},\\
B_{\mu\nu}' & = B_{\mu\nu} - \frac{B_{\mu y} g_{\nu y}-g_{\mu y} B_{\nu y}}{g_{yy}},\\
B_{\mu y}' & = \frac{g_{\mu y}}{g_{yy}},\\
e^{2\phi'} & = \frac{e^{2\phi}}{g_{yy}},
\end{align}
and for the RR-fields:
\begin{align} 
(C^{(n)})' &= (C^{(n)})'_{\textrm{part 1}} + (C^{(n)})_{\textrm{part 2}},\\
(C^{(n)})_{\textrm{part 1}}' &= \left[C^{(n-1)} - \left(C^{(n-1)}_{\mu\cdots \nu y} dx^{\mu}\wedge\cdots dx^{\nu}\right)\wedge \left(\frac{g_{\rho y}}{g_{yy}}dx^{\rho}\right)\right]\wedge dy,\\
 (C^{(n)})_{\textrm{part 2}}' &= C^{(n+1)}_{\mu\cdots \nu y} dx^{\mu}\wedge\cdots dx^{\nu} + \left(C^{(n-1)}_{\mu\cdots \nu} dx^{\mu}\wedge\cdots dx^{\nu}\right)\wedge \left(B_{\rho y} dx^{\rho}\right)\nn
& + \left(C^{(n-1)}_{\mu\cdots \nu y} dx^{\mu}\wedge\cdots dx^{\nu}\right)\wedge \left(B_{\rho y} dx^{\rho}\right) \wedge \left(\frac{g_{\sigma y}}{g_{yy}}dx^{\sigma}\right).\end{align}

Under T-duality on a compact direction $y$ of length $2\pi R_y$, the radius $R_y$ as well as the string coupling $g_s$ change as:
\begin{align}
\label{app:Tdualityradius} R_y &\rightarrow \frac{\alpha'}{R_y}, \\
\label{app:Tdualitygs} g_s & \rightarrow g_s\frac{l_s}{R_y}.
\end{align}

\subsubsection{S-duality}
An S-duality transformation in IIB in our conventions (in the string frame) is given by~\cite{Ortin:2004ms}:
\begin{align}
\phi' &= -\phi,\\
g_{\mu\nu}' & =  e^{-\phi} g_{\mu\nu},\\
B_2' &= -C_2,\\
C_2' &= B_2,\\
\label{app:SdualityC4} C_4' &= C_4 + C_2\wedge B_2.
\end{align}
This is, of course, just one element of the full $SL(2,\mathbb{R})$ symmetry of IIB supergravity.
 
 Note that thus, with these conventions, $C_4$ is \emph{not} invariant under S-duality; the extra term $C_2\wedge B_2$ is absolutely crucial for consistency.\footnote{Of course, if one uses different conventions where e.g. $F_5 = dC_4 - \frac12 dC_2\wedge B_2 + \frac12 C_2\wedge H_3$ such as in~\cite{Dhar:1999ax}, then $C_4$ \emph{is} invariant under S-duality.}

Under S-duality, the string coupling $g_s$ and the string length $l_s=\sqrt{\alpha'}$ change as:
\begin{align}
\label{app:Sdualityls} \alpha'& \rightarrow \alpha' g_s, \\
\label{app:Sdualitygs} g_s &\rightarrow 1/g_s.
\end{align}

\subsection{Units}\label{sec:appunits}
We follow the conventions used in~\cite{Chowdhury:2011qu}, which in turn mainly follows the conventions used in~\cite{Bena:2008dw}.

Newton's constant in $D$ dimensions is given by:
\be G_D = (2\pi)^{D-3} l_D^{D-2}.\ee
The M-theory Planck length is related to the 10D string coupling constant and string length by:
\be l_{11} = g_s^{1/3} l_s,\ee
while the M-theory radius of compactification is:
\be R_{11} = g_s l_s.\ee
The mass of an M2-brane wrapping two compact directions $x,y$ is given by:
\be m_{x,y} = \frac{R_x R_y}{l_{11}^3}.\ee
In a $T^6$ compactification of M-theory, we have an effective five dimensional Newton's constant given by:
\be G_5 = \frac{G_{11}}{vol(T^6)} = \frac{\pi}{4} \frac{g_s^2 l_s^8}{R_z R_1 R_2 R_3 R_4},\ee
where one of the $S^1$'s of the $T^6$ is the M-theory circle (thus the radius is $R_{11}$) and the other five directions are parametrized by our coordinates $z,y_1,y_2,y_3,y_4$.

We will use the following conventions in the final F1-D2-D2 frame (see appendix \ref{sec:appdualities}):
\be \label{appunits:allRequal} R_1 = R_2 = R_3 = R_4 = R_z=R_{11}.\ee
In this frame, we also use the convention:
\be \label{appunits:fixunit} m_{y_4,R_{11}} = m_{y_2,y_3}=m_{z,y_1} = 1,\ee
which implies:
\be \label{appunits:G5} G_5 = \frac{\pi}{4}.\ee
The equality (\ref{appunits:fixunit}) is not dimensionally correct, but should be thought of as fixing the unit of length.

Two interesting implications that can be derived from the unit convention (\ref{appunits:fixunit}) are:
\be \label{appunits:relations} l_s = g_s, \qquad R_{z; 1; 2; 3;4} = l_s^2.\ee

\section{Page Charges}\label{app:page}
The presence of Chern-Simons terms in the 10D supergravity actions effectively ``mix'' the different gauge fields, leading to unintuitive behavior of the charges and currents associated to these fields. It turns out that one can define three notions of charge~\cite{Marolf:2000cb}: Maxwell charge, brane-source charge, and Page charge. We will not discuss Maxwell charge here. Brane-source charge corresponds to the naive definition of charge as the violation of the Bianchi identity for the dual potential. However, brane source charge is not conserved or quantized, so it is not expected to represent physical D-branes (the number of which is certainly a conserved and quantized quantity). In fact, the Page D-brane charge can be argued to be the physical one, as it is conserved and quantized.

In this appendix, we will discuss brane source and especially Page charges for D-branes in IIA and IIB theory and for F1 branes (dissolved in D-branes) in IIA theory. A comprehensive discussion of D-brane charges was recently given in~\cite{deBoer:2012ma}, but to our knowledge no equally extensive discussion of F1 charges is present in the literature. Also, we will take a moment to discuss the interplay between Page charges and large gauge transformations, since this is important for the main body of this article.

In this entire appendix, we will not include the effects of NS5-branes, since we do not deal with them in our main analysis. A discussion of D-brane charges in the presence of NS5-branes is given in~\cite{deBoer:2012ma}; considering F1-charges in the presence of NS5-branes is a topic for future investigation. We will also only consider F1-charge that is dissolved in D-branes and not stacks of F1-strings by themselves.

\subsection{D-brane Page Charges}
A comprehensive and clear discussion of D-brane brane-source and Page charges was recently given in~\cite{deBoer:2012ma}. Here we will briefly mention the definitions and formulas most relevant for our article, in our conventions as detailed above in appendix \ref{app:conventions}.

\subsubsection{Bulk Charges}
A natural but naive definition of D-brane currents is given by the so-called brane source current, which is given by the violation of the Bianchi identities (\ref{app:IIAEOM}). This corresponds (up to a sign) with the functional derivative with respect to $C_p$ of the D-brane action (\ref{appconv:Dbrane}). The resulting definition is, in type IIA and IIB:
\be \label{app:BSdef} dF_p + H_3\wedge F_{p-2} = *j_{D(8-p)}^{BS}.\ee
However, even though this current is in principle localized (as it comes from a source D-brane as in (\ref{appconv:Dbrane})) and manifestly gauge-invariant, it is not conserved. We have:
\be d*j_{Dp}^{BS} = -H_3\wedge *j_{D(p+2)}^{BS}.\ee
The Page D-brane current $j^P_{Dp}$ must be a localized \emph{and} conserved current. We can define the Page current as:
\begin{align}
*j^P_{Dp} &= e^B *j^{BS}_{Dq} && \left(\equiv *j^{BS}_{Dp} + B\wedge *j^{BS}_{D(p+2)}+\cdots\right)  ,\\
\label{app:DbranePagedef} &= d(e^B F) &&\left(\equiv  d\left(F_{8-p} + B\wedge F_{6-p} + \cdot\right)\right) ,
\end{align}
which is manifestly conserved due to the second line (since $dH_3=d^2F_p=0$), and manifestly localized due to the first (since $*j^{BS}$ is localized).

This final expression also makes it easy to calculate the Page charge $Q^P_{Dp}$ contained in a volume $V$, since:
\be Q^P_{Dp} = \int_V *j_{Dp}^P = \int_{\partial V} e^B F, \ee
i.e. it can always be calculated by a surface integral over the boundary $\partial V$.

\subsubsection{Worldvolume Page Charges}
We will illustrate the difference between brane-source and Page charges for a D-brane probe, given by the worldvolume action (\ref{appconv:Dbrane}). For definiteness, we will focus on the case of a D4-brane in IIA, so the relevant WZ part of the action is given by (with $\mathcal{M}_5$ the worldvolume of the brane):
\be S_{WZ}^{D4} = \int_{\mathcal{M}_5} \left(C_5 + (B+F)\wedge C_3 + \frac12 (B+F)\wedge (B+F)\wedge C_1\right). \ee
The brane-source currents, as mentioned above, are (up to a sign) just the derivatives of this action to the potentials $C_p$, so we get:
\begin{align}
 *j_{D4}^{BS} &= +\int_{\mathcal{M}_5} *1,\\
 *j_{D2}^{BS} &= -\int_{\mathcal{M}_5} B+F,\\
 *j_{D0}^{BS} &= +\int_{\mathcal{M}_5} \frac12 (B+F)\wedge (B+F),
\end{align}
while the Page currents are easily found to be given by:
\begin{align}
 *j_{D4}^{P} &= +\int_{\mathcal{M}_5} *1,\\
 *j_{D2}^{P} &= -\int_{\mathcal{M}_5} F,\\
 *j_{D0}^{P} &= +\int_{\mathcal{M}_5} \frac12 F\wedge F.
\end{align}
This fits nicely in with the well-known fact that the background value of $B$ does not have to be quantized, while of course the magnetic field $F$ on a brane is quantized. Thus the Page charge is quantized, as it should be if it is to represent the ``physical'' D-brane charge.

\subsection{F1 Page Charge in IIA} \label{app:pagef1}
For our purposes, it is also important to understand the notion of Page charge for F1 strings. We present a derivation here, since to our knowledge no such discussion is present in the literature yet. (The discussion of~\cite{deBoer:2012ma} only deals with D-brane and NS5-brane charges.) We will focus on F1-charges in type IIA induced by D-brane sources for definitiveness, but the story for IIB should be analogous. Note that we (as above) do not include NS5-brane sources in our discussion; nor do we consider (stacks of) fundamental F1-strings in themselves (i.e. strings that are not dissolved in D-branes).

To discuss F1-charge, we first must derive the equation of motion for $B_2$. We start from the democratic formulation of the bulk action (\ref{appconv:IIAdemocratic}), so that the variation w.r.t. $B_2$ is:
\begin{align}
2\kappa_0^2 \delta S_{IIA} &=\int \left\{ e^{-2\Phi}*H_3\wedge d\delta B_2\right.\\
& \left. +\frac12 *F_4\wedge d\delta B_2\wedge C_1 + \frac12 *F_6\wedge d\delta B_2\wedge C_3 + \frac12 *F_8\wedge d\delta B_2\wedge C_5\right\},\\
& = \int \left\{ -d(e^{-2\Phi}*H_3)\wedge \delta B_2\right.\\
& \left. -\frac12 d(*F_4\wedge C_1)\wedge \delta B_2 - \frac12 d(*F_6\wedge C_3)\wedge \delta B_2 - \frac12 d(*F_8\wedge C_5)\wedge \delta B_2\right\}.
\end{align}
In the presence of D-brane sources, there is also a variation of the DBI+WZ source action to be taken into account:
\be  \label{apppagef1:varsources} 2\kappa_0^2 \delta S_{sources} = 2\kappa_0^2\frac{\delta S_{DBI}}{\delta B}\delta B + \frac12 C\wedge \left(d*F-*F\wedge H\right)\wedge \delta B .\ee
The factor of $1/2$ in the terms coming from the WZ action is because we are working in the democratic formulation.

The equation of motion will thus have contributions proportional to $j^{BS}_{Dp}$ (or equivalently, $\left(d*F-*F\wedge H\right)$) coming from the variations of both the bulk and source terms. In fact, these two contributions cancel each other exactly, leaving us with the equation of motion:
\be d(e^{-2\Phi}*H_3) -F_2\wedge F_6 +\frac12 F_4\wedge F_4 = \frac{\delta S_{DBI}}{\delta B}.\ee

Brane source charge is expected to be localized and gauge invariant; this expression is manifestly both, so it is natural to define:
\be *j^{BS}_{F1} = \frac{\delta S_{DBI}}{\delta B}.\ee
However, this is not conserved, as we can see by direct calculation:
\begin{align}
 d*j^{BS}_{F1} &= -dF_2\wedge F_6-F_2\wedge dF_6 + dF_4\wedge F_4\\
& = -*j^{BS}_{D6}\wedge F_6 -F_2\wedge *j^{BS}_{D2} +*j_{D4}^{BS}\wedge F_4.
\end{align}
So we must define the conserved Page charge as:\footnote{We assume there are no D8-branes present.}
\be *j^{P}_{F1} = *j^{BS}_{F1} + C_5\wedge *j^{BS}_{D6} - C_3\wedge *j_{D4}^{BS} + C_1\wedge *j^{BS}_{D2},\ee
which gives us easily:
\be d*j^P_{F1} = 0.\ee
Note that, for a D-brane source, this expression is exactly:
\be \label{apppagef1:pagef1current} *j^P_{F1} = \frac{\delta (S_{DBI} + S_{WZ})}{\delta B_2}.\ee
Note that, as opposed to the expression (\ref{apppagef1:varsources}) where there is a factor of 1/2 multiplying the variation of $S_{WZ}$, we observe that in the derived expression (\ref{apppagef1:pagef1current}) there is no such factor of 1/2.

\subsection{Large Gauge Transformations of $B$}\label{sec:apppagelargegauge}
Since the Page current is expected to be the physical current for D-branes, this means that in addition to being localized and conserved, the Page charge should be \emph{quantized} as well. However, from the definition (\ref{app:DbranePagedef}) of the Page current, it is apparently not even gauge invariant, which raises the question of how it could be quantized at all.

Consider first a small gauge transformation $\delta B_2 = d\Lambda$. If we calculate the resulting gauge variation of the Page charge contained in a volume $V$, we get:
\be \delta Q^P = \int_{\partial V} d\Lambda\wedge e^B F = -\int_{\partial V} \Lambda\wedge d(e^B F) = -\int_{\partial V} \Lambda\wedge *j^P ,\ee
so if the volume completely encloses the (localized) Page currents present, the Page charge in this volume is invariant under small gauge transformations~\cite{Marolf:2000cb}.

However, the Page charge will in general not be invariant under large gauge transformations, under which it is possible that
\be \label{app:Blargegauge} \delta B = C d\alpha\wedge d\beta,\ee
 where $\alpha,\beta$ are compact directions and $C$ is a constant. Now we see how it is possible to retain quantization of the Page charge: not every value of $C$ in (\ref{app:Blargegauge}) is obtainable by a large gauge transformation; the possible values of $C$ are \emph{quantized}!\footnote{That such large gauge transformations are quantized is a fact which can easily be seen for a one-form, but to derive this for a two-form more mathematical structures such as gerbes must be used.}

We can also use this reasoning in reverse, and derive the quantized allowed values of $C$ for $\delta B$ assuming that Page charge is quantized. Let us assume for simplicity that we have a D2-brane which wraps the two compact directions $\alpha,\beta$ of radii $R_{\alpha},R_{\beta}$. Initially, we start with the field on the brane as well as $B_2$ set to zero, $F_{\alpha\beta}=B_{\alpha\beta}=0$. Then, we perform a large gauge transformation as above, so that we now have $B_{\alpha\beta} = C$. We know that $F+B$ must be gauge invariant, so now we must have $F_{\alpha\beta} = -C$. However, this induces a non-zero D0-brane Page charge:
\be *j^{D0}_P = -T_{D2}\int d\alpha d\beta F_{\alpha\beta} = T_{D2} C (2\pi R_{\alpha}) (2\pi R_{\beta}).\ee
Using the quantization of Page charge, this last expression must be equal to $N T_{D0}$ with $N$ an integer. So we see that:
\be \label{app:Clargegauge} C = N \frac{\alpha'}{R_{\alpha} R_{\beta}}. \ee
This is, then, the allowed values of large gauge transformations which change $B_{\alpha\beta}$ by a constant.

In the duality chain described below, we add the constant piece $b_{13} dy^1\wedge dy^3 + b_{24}dy^2\wedge dy^4$ to the $B_2$ field in the D1-D5-P frame at the end of the chain (\ref{app:dualitychainfull2}). This can be seen as a large gauge transformation of the $B$ field for certain values of $b_{13}$, corresponding to (\ref{app:Clargegauge}). We can follow this value through the final duality chain (\ref{app:dualitychainfull3}) to see what it will be in the final F1-D2-D2 frame, by using the duality rules (\ref{app:Tdualityradius})-(\ref{app:Tdualitygs}) and (\ref{app:Sdualityls})-(\ref{app:Sdualitygs}):
\be b_{13} = N \frac{\alpha'}{R_1 R_3} \xrightarrow[T_{y_4,z}]{} N \frac{\alpha'}{R_1 R_3} \xrightarrow[S]{} N\frac{\alpha' g_s}{R_1 R_3} \xrightarrow[T_{y_1}]{} N\frac{g_s l_s}{R_3} .\ee
In our unit conventions in the final F1-D2-D2 frame (see (\ref{appunits:relations}) in appendix \ref{sec:appunits}), $g_s l_s/R_3$ is exactly 1, so that in our conventions the values that $b_{13}$ can take (if it is to be seen as a large gauge transformation) are precisely the integers.

\section{Duality Calculations}\label{sec:appdualities}
In this appendix, we explain in more detail the duality transformations used in this article.

We choose to start from the three-charge non-extremal, rotating 5D black hole in an M-theory frame. There are three stacks of M2-branes wrapping perpendicular cycles on the compact $T^6$. The system is given by:
\begin{align}
ds_{11}^2 &= -(H_1 H_2 H_3)^{-2/3}H_m(dt+k)^2 + (H_1 H_2 H_3)^{1/3} ds_4^2\nn
& + \frac{(H_1 H_3)^{1/3}}{H_2^{2/3}}\left(dy_1^2+dy_2^2\right) + \frac{(H_2 H_3)^{1/3}}{H_1^{2/3}}\left(dy_3^2+dy_4^2\right) + \frac{(H_1 H_2)^{1/3}}{H_3^{2/3}}\left(dz^2+dx_{11}^2\right),\\
\mathcal{A}_3 & = A_2\wedge dy_1\wedge dy_2 + A_1\wedge dy_3\wedge dy_4 + A_3\wedge dz\wedge dx_{11},
\end{align}
where we have used the definition:
\begin{equation}
 A_i = \frac{c_i}{s_i}H_i^{-1}(dt+k)-\frac{c_i}{s_i} dt +B_i,
\end{equation}
in addition to the other definitions of building blocks in section \ref{sec:modinitial}.

Starting from this M-theory system, we can reduce on the $S^1$ parametrized by $x^{11}$ to get to a D2-D2-F1 frame. To get to the D1-D5-P frame, we simply do three T-dualities along $z,y_3,y_4$. In this frame, $\delta_1$ is the parameter that determines the D1-charge (in the sense of (\ref{mod:initialQi})), $\delta_2$ determines the D5-charge, and $\delta_3$ the P-charge, so this is precisely the initial D1-D5-P system quoted in (\ref{mod:metricinitial})-(\ref{mod:B2C4initial}). So far this is the duality sequence (\ref{app:dualitychainfull1}).

To turn on the moduli as per the procedure of~\cite{Maldacena:1999mh,Dhar:1999ax,Lunin:2003gw}, we continue now on the second sequence depicted in (\ref{app:dualitychainfull2}). We first dualize back from the D1-D5-P frame to a D3-D3-P frame by T-dualities along $y_3, y_4$. The two D3-stacks wrap the directions $(z,y_1,y_2)$ and $(z,y_3,y_4)$, respectively. In this frame, we can rotate by an angle $\xi$ in the $(y_1,y_3)$ plane and by an angle $\omega$ in the $(y_2,y_4)$ plane. This means our original D3-brane stacks are now at an angle in these planes; this will change the behavior of the system under T-duality on these directions. For example, if we now perform a T-duality along $y_3$, we will obtain D2-branes along $(z,y_4)$ and D4-branes along $(z,y_1,y_2,y_3)$ as before, but we will also have new D2-branes along $(z,y_2)$ and new D4-branes along $(z,y_1,y_3,y_4)$. So when, after the two rotations described above, we again do T-dualities along $y_3, y_4$, we will be in a D1-D5-P frame with extra D3-branes wrapping $(z,y_1,y_3)$ and $(z,y_2,y_4)$. In this final frame we also just add a constant piece $b_{13} dy_1\wedge dy_3 + b_{24} dy_2\wedge dy_4$ to the NS 2-form. The end result is the D1-D5-P(-D3-D3) system described by (\ref{mod:metricwmod})-(\ref{mod:C6wmod}).

However, in this frame the calculation of the supertube Lagrangian and Hamiltonian are difficult since it would be a KK-dipole with D1 and D5 charges as depicted in (\ref{app:dualitychainfull2}). To be able to use a standard D-brane DBI+WZ action in calculating the probe supertube Hamiltonian, we do the final sequence of dualities (\ref{app:dualitychainfull3}) on this system (\ref{mod:metricwmod})-(\ref{mod:C6wmod}), to obtain a more friendly frame for calculating the worldvolume action of the probe. By performing the chain of dualities $T_{y_4,z},S, T_{y_1}$, we transform the D1-D5-P(-D3-D3) system into a F1-D2-D2(-F1-D2) system, where the initial D1-charge is now F1-charge (wrapping $y_4$), the initial D5-charge is now a D2-charge (wrapping $y_2,y_3$), the initial P-charge has become another D2-charge (wrapping $z,y_1$), and the two initial stacks of D3's have become additional F1 and D2 charge (along $y_2$ and $y_3,y_4$, respectively). In this final frame, the supertube has become a D4-brane dipole wrapping $y_2,y_3,y_4$ and a contractible cycle in the non-compact space given by (\ref{mod:probecycle}).

The entire sequence of dualities has been mapped out in detail in (\ref{app:dualitychainfull1})-(\ref{app:dualitychainfull3}), keeping track of all the charges along the way. Note that $\alpha$ denotes the cycle in the non-compact space that the probe dipole wraps, see (\ref{mod:probecycle}).

\begin{align}
 &\begin{cases}  M2(y_3,y_4)\\ M2(y_1,y_2)\\ M2(z,x^{11}) \end{cases} &\xrightarrow[S^1(x^{11})]{} & \begin{cases} D2(y_3,y_4)\\ D2(y_1,y_2)\\ F1(z) \end{cases}  &\xrightarrow[T_{z,y_3,y_4}]{} & \begin{cases} D1(z)\\ D5(z,y_1,y_2,y_3,y_4)\\ P(z) \end{cases}\nn
\label{app:dualitychainfull1}\textrm{Probe:} & \quad M5(y_1,y_2,y_3,y_4,\alpha) && \quad NS5(y_1,y_2,y_3,y_4,\alpha) && \quad KK(y_1,y_2,y_3,y_4,\alpha;z)\\ \nn
\xrightarrow[T_{y_3,y_4}]{} & \begin{cases} D3(z,y_3,y_4)\\ D3(z,y_1,y_2)\\ P(z) \end{cases}  & \xrightarrow[\xi(y_1,y_3);\ \omega(y_2,y_4)]{} & \begin{cases} D3(z,y_3,y_4)\\ D3(z,y_1,y_2)\\ P(z)\\ D3(z,y_2,y_3)\\ D3(z,y_1,y_4) \end{cases} & \xrightarrow[T_{y_3,y_4}]{} & \begin{cases} D1(z)\\ D5(z,y_1,y_2,y_3,y_4)\\ P(z)\\ D3(z,y_2,y_4)\\ D3(z,y_1,y_3) \end{cases}\nn
\label{app:dualitychainfull2}\textrm{Probe:} & \quad KK(y_1,y_2,y_3,y_4,\alpha;z) && \quad KK(y_1,y_2,y_3,y_4,\alpha;z) && \quad KK(y_1,y_2,y_3,y_4,\alpha;z)\\ \nn
\xrightarrow[T_{y_4,z}]{} & \begin{cases} D1(y_4)\\ D3(y_1,y_2,y_3)\\ F1(z)\\ D1(y_2)\\ D3(y_1,y_3,y_4) \end{cases} & \xrightarrow[S]{} & \begin{cases} F1(y_4)\\ D3(y_1,y_2,y_3)\\ D1(z)\\ F1(y_2)\\ D3(y_1,y_3,y_4) \end{cases}  & \xrightarrow[T_{y_1}]{} & \begin{cases} F1(y_4)\\ D2(y_2,y_3)\\ D2(z,y_1)\\ F1(y_2)\\ D2(y_3,y_4) \end{cases}\nn
\label{app:dualitychainfull3}\textrm{Probe:} & \quad NS5(y_1,y_2,y_3,y_4,\alpha) && \quad D5(y_1,y_2,y_3,y_4,\alpha) && \quad D4(y_2,y_3,y_4,\alpha)
\end{align}

\section{The Probe Hamiltonian}
In this appendix, we discuss the calculations involving finding the probe Hamiltonian and its decoupling limit, and investigating some of its properties.

\subsection{Calculating the Hamiltonian}\label{sec:appham}
In the final F1-D2-D2 frame discussed in appendix \ref{sec:appdualities}, the probe we are interested in is a D4-brane dipole wrapping the compact directions $y_2,y_3,y_4$ as well as one contractible cycle in the non-compact space. We define the embedding coordinates as:
\begin{align}
 \xi^0 &= t, & \xi^1 &= y_4, & \xi^3&= y_2, & \xi^4 &= y_3,
\end{align}
and the cycle in the 5D non-compact space is determined by two embedding constants $b_1,b_2$ as:
\begin{align}
 \psi&=b_1 \xi^2,& \phi&= b_2\xi^2 .
\end{align}
The reason why we choose our embedding coordinates in this perhaps peculiar way is to make contact with~\cite{Chowdhury:2011qu}, where $\xi^1$ is also the direction along which the original F1-charge runs.

The D4-brane action is (all relevant quantities are pulled back on the worldvolume):
\begin{align}
S_{D4} &= S_{DBI} + S_{WZ},\\
S_{DBI} &= -|N_{D4}| T_{D4}\int d^5\xi e^{-\phi}\sqrt{-\det\left(g+B+F\right)},\\
S_{WZ} &= N_{D4} T_{D4} \int d^5\xi \left( C_5 + C_3\wedge(B+F) + \frac12 C_1\wedge(B+F)\wedge(B+F)\right).
\end{align}
In pulling back quantities, one must be careful about minus signs. For example:
\begin{equation}
 (C_5)_{01234} = -b_2 (C_5)_{t\psi y_2 y_3 y_4} - b_1 (C_5)_{t\phi y_2 y_3 y_4}.
\end{equation}
As for the gauge field on the brane, we will turn on an electric field $\epsilon$ and a magnetic field $\beta$, which resp. determine the F1 and D2 charges along the same directions as the F1 and D2 charges of the background. In addition, we will also turn on a second electric field $\epsilon_3$ and a second magnetic field $\beta_3$, which correspond to the extra F1 and D2 charges of the background as discussed above (the subscript $3$ on these fields is to remind us that they correspond to the two D3-charges in the D1-D5-P frame). In total, we have:
\be
F = \epsilon \ d\xi^0\wedge d\xi^1 + \beta\ d\xi^1\wedge d\xi^2 + \epsilon_3\ d\xi^0\wedge d\xi^3 + \beta_3\ d\xi^2\wedge d\xi^3.\ee
For shorthand, we can define the shifted fields:
\begin{align}
 \tilde{\epsilon} &= \epsilon + B_{01},& \tilde{\epsilon_3} &= \epsilon_3 + B_{03}, & \tilde{\beta} &= \beta + B_{13}, & \tilde{\beta_3} &= \beta_3 + B_{23}.
\end{align}
Using this, the Lagrangian can be written as:
\begin{align}
 \frac{L}{m_1 m_2 N_{D4}} &= -sgn(N_{D4}) e^{-\phi}\sqrt{-g_{44}}\left\{g_{33}\left[-g_{02}^2g_{11}+g_{00}\left(g_{11}g_{22}+\tilde{\beta}^2\right)+g_{02}\tilde{\epsilon}\tilde{\beta}+g_{22}\tilde{\epsilon}^2\right] \right.\nn
 &\left. + \tilde{\epsilon_3}\left[\tilde{\beta_3}(-g_{02}g_{11}+\tilde{\epsilon}\tilde{\beta})-g_{22}(g_{13}\tilde{\epsilon}-g_{11}\tilde{\epsilon_3})+\tilde{\beta}(-g_{02}g_{13}+\tilde{\beta}\tilde{\epsilon_3})\right] \right.\nn
 & \left.  +g_{13}\left[\tilde{\beta_3}(g_{00}\tilde{\beta}+g_{02}\tilde{\epsilon})+g_{02}(g_{02}g_{13}-\tilde{\beta}\tilde{\epsilon_3})-g_{22}(g_{00}g_{13}+\tilde{\epsilon}\tilde{\epsilon_3})\right] \right.\nn
&\left. +\tilde{\beta_3}\left[\tilde{\beta_3}(g_{00}g_{11}+\tilde{\epsilon}^2)+g_{02}(g_{13}\tilde{\epsilon}-g_{11}\tilde{\epsilon_3})+\tilde{\beta}(g_{00}g_{13}+\tilde{\epsilon}\tilde{\epsilon_3})\right]\right\}^{1/2}\nn
&+(C_5)_{01234} + \left((C_3)_{014}\tilde{\beta_3}+(C_3)_{034}\tilde{\beta}+(C_3)_{124}\tilde{\epsilon_3}+(C_3)_{234}\tilde{\epsilon}\right) + (C_1)_4 \left(\tilde{\epsilon}\tilde{\beta_3} + \tilde{\epsilon_3}\tilde{\beta} \right).
\end{align}
Note that:
\be m_1 m_2 = T_{D4} \int d\xi^1 d\xi^2 d\xi^3 d\xi^4 = T_{D4} (2\pi R_4) (2\pi) (2\pi R_2) (2\pi R_3).\ee

To calculate the Hamiltonian given the Lagrangian, we must perform a double Legendre transform with respect to the electric fields $\epsilon,\epsilon_3$. We define the conserved charges (in units where $G_5=\pi/4$):
\begin{align}
\label{appham:q1} q_1 &= \frac{\partial L}{\partial \epsilon},\\
\label{appham:q3}q_3 &= \frac{\partial L}{\partial \epsilon_3}.
\end{align}
Note that, as per the discussion in appendix \ref{app:pagef1}, $q_1$ and $q_3$ correspond to the quantized, physical Page F1 charges induced on the D4-brane.

In principle, we need to solve the equations (\ref{appham:q1}) and (\ref{appham:q3}) simultaneously to get the functions $\epsilon(q_1,q_3), \epsilon_3(q_1,q_3)$. Then, we need to enter these expressions in the Legendre transform:
\be H(q_1,q_3) = q_1\epsilon(q_1,q_3) + q_3\epsilon_3(q_1,q_3) - L\left[\epsilon(q_1,q_3),\epsilon_3(q_1,q_3)\right].\ee
However, solving the two equations simultaneously appears to be impossible analytically, so we must result to numerics in calculating $\epsilon(q_1,q_3), \epsilon_3(q_1,q_3)$ and the resulting Hamiltonian.

The final Hamiltonian also contains the two magnetic fields, which correspond to (the absolute value of) Page D2 charges (see appendix \ref{app:page}) which we define as (again, in units where $G_5=\pi/4$):
\bea
q_2 &=& d\ \beta,\\
q_3' &=& d\ \beta_3,
\eea
where $d$ is the D4-dipole charge of the probe.

So in the end, the Hamiltonian should be viewed as a function of the four dissolved charges $q_1, q_2, q_3, q_3'$ and the D4-dipole charge $d$, as well as the embedding coordinates $b_1, b_2$. We always take $\theta=0$ but we will vary $r$, so we can view the Hamiltonian as being a function of $r$ as well. So in the end our Hamiltonian depends on 8 parameters:
\begin{equation}
 \mathcal{H} = \mathcal{H}(r, b_1, b_2,d,q_1,q_2,q_3,q_3'),
\end{equation}
which is exactly (\ref{mod:endham}).

\subsection{The Decoupling Limit Hamiltonian}\label{sec:apphamdecouple}
The decoupling limit of the D1/D5 geometry is taken by assuming that the $S^1$ direction that the D1 and D5 branes share is large compared to the other compact $T^4$ directions that the D5-branes wrap. This is achieved by assuming:
\be Q_{D1}^P,Q_{D5}^P \gg Q_3,m,a_1^2, a_2^2,\ee
and considering the region:
\be r^2\ll Q_{D1}^P, Q_{D5}^P.\ee
We state these conditions here without further explanation; the reader is referred to e.g. section 4 of~\cite{Chowdhury:2011qu} for a discussion on the decoupling limit of our system.

Practically, we can implement the decoupling limit of the Hamiltonian of the tube by setting:\footnote{Note that, at finite moduli, $N_{1,2}$ are not the Page D1/D5 charges of the system, but rather these parameters are connected to the Page charges in the way that we are familiar with from (\ref{mod:D5chargewmod})-(\ref{mod:D1chargewmod}).}
\begin{align}
 H_{1,2} &= \frac{1}{\delta}\frac{N_{1,2}}{r^2},& Q_{1,2} &= \frac{1}{\delta}N_{1,2},\nn
\label{appham:decouplinglim} q_{1,2}&=\frac{1}{\delta}N^{tube}_{1,2},& \frac{4 G_5}{\pi} &= \frac{1}{\delta}.
  \end{align}
We will always work with $q_3=q_3'=0$ when considering the decoupling limit. We then expand the Hamiltonian to first order in $\delta$:
\be \mathcal{H} = \mathcal{H}_0 +  \mathcal{H}_1\delta+O(\delta^2),\ee
and then the decoupling limit Hamiltonian is defined as:
\be \mathscr{H} = \mathcal{H}_1.\ee

Unfortunately, since we don't have an analytic expression for $\mathcal{H}$ as discussed in the previous section, we cannot just implement (\ref{appham:decouplinglim}) immediately at the level of the Hamiltonian. Instead, we need to work with the Lagrangian. This has the added difficulty that it depends on the electric/magnetic fields $\epsilon,\epsilon_3,\beta,\beta_3$, of which the $\delta$ dependence must be figured out before we can consistently take the series expansion of the Lagrangian in $\delta$. We will sketch the procedure to do this here, but we omit the actual calculations, which involve the enormous expressions of the explicit form of the Lagrangian and were all performed using Mathematica.

The $\delta$ dependence of the magnetic fields $\beta,\beta_3$ is easy to figure out: they are independent of $\delta$ (i.e. $O(\delta^0)$). This is because e.g. $q_2\sim \delta^{-1}$ as mentioned above, and $q_2 = d\ \beta$, where the dipole charge $d$ contains a factor of $\delta^{-1}$.

The $\delta$ dependence of the electric fields $\epsilon,\epsilon_3$ is not so straightforward, but still quite possible to determine. We can use the zero moduli Hamiltonian, of which the explicit analytic expression was found in~\cite{Chowdhury:2011qu}. This teaches us that the electric fields have a series expansion in $\delta$, e.g.:
\be \epsilon = \epsilon^{(0)} + \epsilon^{(1)}\delta + \epsilon^{(2)}\delta^2 + O(\delta^3) .\ee
The lowest order term $\epsilon^{(0)}$ in this expansion is independent of the corresponding F1 charges ($q_1,q_3$) and is determined solely by demanding that the expression $\partial L/\partial \epsilon$ does not contain an imaginary part. The $\epsilon^{(1)}$ term depends on and determines the F1 charges, and all higher order terms are also fixed by $\epsilon^{(1)}$. For determining the decoupling limit Hamiltonian, we only need the coefficients $\epsilon^{(0)},\epsilon^{(1)}$. In fact, we have (remembering that $q_3=0$):
\be \label{appham:decouplingham} \mathscr{H} = \frac{\partial L}{\partial \epsilon} \epsilon^{(1)} - L_1,\ee
where $L_1$ is the $O(\delta^1)$ term in the $\delta$-expansion of the Lagrangian, $L = L_0 +  L_1\delta + O(\delta^2)$.

Proving the invariance of $\mathscr{H}$ under $\xi\rightarrow \pi/2-\xi$ (see section \ref{sec:extmodmoduli}) now comes down to proving the invariance of each term in (\ref{appham:decouplingham}). Note that under $\xi\rightarrow \pi/2-\xi$, we also need to take $N_1\rightarrow N_1\tan^2\xi$ and $N_2\rightarrow N_2\cot^2\xi$ to keep the Page charges $Q_{D1}^P,Q_{D5}^P$ fixed.

 We can first prove by direct calculation that $\partial L/\partial \epsilon$ is invariant under this transformation (up to first order in $\delta$), which implies that also $\epsilon,\epsilon_3$ are invariant. Finally, we use this to find that $L_1$ is invariant as well, thus completing the proof that $\mathscr{H}$ is invariant.

 Finally, to prove the correct scaling of $\mathscr{H}$ with $\eta$ (i.e. its scaling with the string coupling; see section \ref{sec:extmodqual}), a similar procedure can be used. First, we note that $\partial L/\partial \epsilon$ must scale as $\eta^1$; this implies that the relevant electric field components scale as $\epsilon^{(0)}\rightarrow \epsilon^{(0)}$ and $\epsilon^{(1)}\rightarrow\eta\,\epsilon^{(1)}$. The magnetic field obviously scales as $\beta\rightarrow\eta\,\beta$. Then, we can use this to prove that the scaling of $L_1$ is $L_1\rightarrow \eta^2\, L_1$. This implies that $\mathscr{H}\rightarrow \eta^2\mathscr{H}$, as expected.

\bibliographystyle{JHEP}
\bibliography{Papers}

\end{document}